\documentclass[sigconf]{acmart}

\let\url\nolinkurl
\usepackage{balance} 
\usepackage{graphicx} 
\usepackage{textcomp} 
\usepackage{xcolor} 
\usepackage{array} 
\usepackage{caption} 
\usepackage{tabularx,ragged2e,booktabs} 
\usepackage{enumitem} 
\usepackage[autostyle=false, style=english]{csquotes}
\MakeOuterQuote{"}
\newcolumntype{Z}{ >{\centering\arraybackslash}X }
\usepackage{listings}
\definecolor{codegreen}{rgb}{0.0,0.0,0.0}
\definecolor{codegray}{rgb}{0.5,0.5,0.5}
\definecolor{codepurple}{rgb}{0.58,0,0.82}
\definecolor{backcolour}{rgb}{0.97, 0.96, 1.0}
\AtBeginDocument{%
  }

\copyrightyear{2025}
\acmYear{2025}
\setcopyright{cc}
\setcctype{by-nc}
\acmConference[Websci '25]{17th ACM Web Science Conference}{May 20--24, 2025}{New Brunswick, NJ, USA}
\acmBooktitle{17th ACM Web Science Conference (Websci '25), May 20--24, 2025, New Brunswick, NJ, USA}
\acmDOI{10.1145/3717867.3717925}
\acmISBN{979-8-4007-1483-2/2025/05}

\begin{document}

\title{Not Here, Go There: Analyzing Redirection Patterns on the Web}\titlenote{This is an extended version of the  paper accepted for publication at the 17th ACM Web Science Conference (Websci '25)}

\author{Kritika Garg}
\orcid{0000-0001-6498-7391}
\affiliation{%
  \institution{Old Dominion University}
  \city{Norfolk}
  \state{VA}
  \country{USA}
}
\email{kgarg001@odu.edu}

\author{Sawood Alam}
\orcid{0000-0002-8267-3326}
\affiliation{%
  \institution{Internet Archive}
  \city{San Francisco}
  \state{CA}
  \country{USA}
}
\email{sawood@archive.org}

\author{Dietrich Ayala}
\orcid{0000-0002-8255-5805}
\affiliation{%
  \institution{Protocol Labs}
  \city{San Francisco}
  \state{CA}
  \country{USA}
}
\email{dietrich@protocol.ai}

\author{Michele C. Weigle}
\orcid{0000-0002-2787-7166}
\affiliation{%
  \institution{Old Dominion University}
  \city{Norfolk}
  \state{VA}
  \country{USA}
}
\email{mweigle@cs.odu.edu}

\author{Michael L. Nelson}
\orcid{0000-0003-3749-8116}
\affiliation{%
  \institution{Old Dominion University}
  \city{Norfolk}
  \state{VA}
  \country{USA}
}
\email{mln@cs.odu.edu}

\renewcommand{\shortauthors}{Garg et al.}

\begin{abstract}
URI redirections are integral to web management, supporting structural changes, SEO optimization, and security. However, their complexities affect usability, SEO performance, and digital preservation. This study analyzed 11 million unique redirecting URIs, following redirections up to 10 hops per URI, to uncover patterns and implications of redirection practices. Our findings revealed that 50\% of the URIs terminated successfully, while 50\% resulted in errors, including 0.06\% exceeding 10 hops. Canonical redirects, such as HTTP to HTTPS transitions, were prevalent, reflecting adherence to SEO best practices. Non-canonical redirects, often involving domain or path changes, highlighted significant web migrations, rebranding, and security risks. Notable patterns included ``sink'' URIs, where multiple redirects converged, ranging from traffic consolidation by global websites to deliberate ``Rickrolling.'' The study also identified 62,000 custom 404 URIs, almost half being soft 404s, which could compromise SEO and user experience. These findings underscore the critical role of URI redirects in shaping the web while exposing challenges such as outdated URIs, server instability, and improper error handling. This research offers a detailed analysis of URI redirection practices, providing insights into their prevalence, types, and outcomes. By examining a large dataset, we highlight inefficiencies in redirection chains and examine patterns such as the use of ``sink'' URIs and custom error pages.  This information can help webmasters, researchers, and digital archivists improve web usability, optimize resource allocation, and safeguard valuable online content.

\end{abstract}

\ccsdesc[500]{Information systems~World Wide Web}

\keywords{Web Science, HTTP Redirection, SEO, Web Archiving}

\maketitle

\section{Introduction}\label{sec:intro}

The web is an ever-evolving ecosystem, with web pages constantly being updated, restructured, or deleted, reflecting the continuous changes on the web~\cite{BernersLee1994}. These ongoing changes present significant challenges for researchers studying web content~\cite{koehler1999webpageconstancy,cho2000webevolution,Fetterly2003,WebScience2012}. A prominent issue stemming from this dynamic nature is link rot, where hyperlinks lead to inaccessible resources, complicating web studies and diminishing the reliability of hyperlinked information~\cite{klein2014scholarly,jones2016scholarly, zittrain2021paper, pew2024}. Web archives like Internet Archive (IA)'s Wayback Machine mitigate these challenges by preserving web snapshots, enabling access to historical content for studying web evolution. This work~\cite{nypwblog2021} is part of our broader research effort examining the evolution of the web, focusing on its changes, emerging features, and the complexities of tracking its perpetually shifting content~\cite{holzmann2016dawn,agata2014lifespan}. For our previous study~\cite{Samplingtechreport:2025,nypwfindingsblog2024}, we analyzed a dataset of 27.3 million URIs and found that 11 million (40\%) were redirecting, prompting us to investigate the patterns and implications of these redirects. Redirects play a critical role in mitigating link rot, preserving search engine optimization (SEO) value, and facilitating seamless content relocation~\cite{rfc7231}. However, they also introduce complexities, such as reduced usability, challenges for SEO performance, and issues in digital preservation, highlighting the need to analyze their broader implications.

In this study, we crawled and analyzed 11 million unique redirecting URIs, encompassing over 2.6 million unique domains. To delve deeper into the complexities of URI redirections and their broader implications, we focused our analysis on the following key aspects:
\begin{itemize}
[noitemsep, topsep=0pt, leftmargin=0.5cm]
    \item \textbf{Prevalence and patterns of URI redirections:} We examined the overall occurrence of redirections, analyzing hop counts, termination statuses, and the distribution of canonical versus non-canonical redirections. Canonical redirects, such as HTTP to HTTPS transitions, enforce a preferred URL version, while non-canonical redirects often involve domain or path changes.
    \item \textbf{Types and frequency of canonical redirections:} Our analysis encompassed redirections from HTTP to HTTPS, WWW to non-WWW, and vice versa, shedding light on the adoption of secure and streamlined URI structures.
    \item \textbf{Impact on URI path depth and domain changes in redirections:} We analyzed non-canonical redirections to assess whether they maintained, simplified, or added complexity to URI structures by examining changes in path depth. Additionally, we categorized these redirections based on their involvement in host, subdomain, or top-level domain (TLD) changes, providing insights into domain-level transitions.
    \item \textbf{Emergence of redirection "sink" URIs:} Our research identified and analyzed "sink" URIs, where multiple source URIs converge, revealing patterns related to consolidation, affiliate marketing, error pages, and pranks.
    \item \textbf{Prevalence of  error pages:} We examine custom 404 error pages alongside soft 404 errors~\cite{bar2004sic}, where incorrect HTTP status codes mask broken links, and discuss their implications for web usability and preservation.
\end{itemize}

This comprehensive examination provides insights into URI redirection patterns and their broader impact on web usability, digital preservation, and cybersecurity. Our findings offer  recommendations for webmasters to optimize redirection strategies, address soft 404 errors, and improve website management. Additionally, our research contributes to a better understanding of redirection practices, which could inform the development of more robust and efficient web crawlers. By addressing the complexities of URI redirections, this study advances our understanding of the evolving web and its implications for researchers, web developers, and digital archivists alike.

\section{Background and Related Work}\label{sec:background}
URI redirection is a critical component of web architecture, facilitating the seamless transfer of users from one web resource to another. As the web grows increasingly complex, understanding the mechanisms, patterns, and implications of URI redirections has become essential for optimizing web performance, enhancing user experience, and ensuring web security \cite{thompson2024}.

Web crawlers, such as the IA’s Heritrix~\cite{Heritrix3Wiki}, play a vital role in capturing web data for archival and research purposes. Heritrix is a widely-used, open-source web crawler designed for web archiving. It systematically navigates through the web by following links and saving web pages, which are then stored in formats like WARC (Web ARChive) files~\cite{iso2017warc}. One of the key parameters in web crawling is the hop count, which refers to the number of links (or hops) the crawler follows from the starting URI (Uniform Resource Identifier)~\cite{rfc3986}. In most practical scenarios, legitimate redirects rarely exceed 10 hops. Following more than 10 redirects often indicates an unusual or problematic situation, such as a redirection loop or a poorly configured web server. Limiting hops helps avoid infinite loops and optimizes crawling. For instance, Googlebot sets a default limit of 5 hops to prevent latency and manage crawl budgets effectively~\cite{MuellerRedirectAdvice}.

A CDX file serves as an index to the WARC files generated by Heritrix, containing metadata about each captured web resource. Figure~\ref{fig:timemap} presents an example snippet of a CDX response from the IA’s CDX server for a specific URI. This CDX response provides a TimeMap, which is a list of archived versions, or mementos, associated with the given URI \cite{memento:rfc}. The CDX index contains SURT (Sort-friendly URI Reordering Transform) which is a method to create a standardized, canonicalized key for URIs within the TimeMap. For example, in Figure~\ref{fig:timemap}, the SURT representation (blue text) canonicalizes various URI variations(red text), such as those with or without \texttt{"www"}, trailing slashes, and different protocol schemes (HTTP vs. HTTPS).

\begin{figure*} 
\begin{lstlisting}[numbers=none, backgroundcolor = \color{white}]
<@\textcolor{blue}{au,com,ecogeneration)/}@> 20220126154911 <@\textcolor{red}{http://ecogeneration.com.au/}@> text/html 301  568
<@\textcolor{blue}{au,com,ecogeneration)/}@> 20220126154912 <@\textcolor{red}{https://ecogeneration.com.au/}@> text/html 301  372
<@\textcolor{blue}{au,com,ecogeneration)/}@> 20220126154913 <@\textcolor{red}{https://www.ecogeneration.com.au/}@> text/html 200  17607
  
\end{lstlisting}
\caption{A snippet of a \texttt{TimeMap} retrieved from the IA's CDX Server. These fields represent the SURT (canonicalized URI) (blue text), the datetime, the original URI (red text), the MIME type of the original document, the HTTP response code, and the length of the response record. }
  \Description{}
\label{fig:timemap}
\end{figure*}

Canonicalized redirects refer to redirecting multiple URIs to a single, authoritative URI, known as the canonical URI~\cite{rfc6596}. This technique is used to prevent issues related to duplicate content, where different URIs lead to the same or very similar content. The primary purpose of canonicalization is to inform search engines which URI should be considered the master copy, ensuring that link equity is consolidated to the canonical URI. These redirects unify various URI versions into a single canonical URI, mapping to the same TimeMap (Figure~\ref{fig:timemap}). Canonical redirects are typically seamless for end users because they occur automatically and transparently during the web browsing experience. When a user requests a URI that is not the canonical version (such as one that includes a "www" subdomain or a trailing slash), the server automatically redirects the user to the canonical version of the URI. Table~\ref{tab:Ctypes} shows various examples of canonical redirects.

\begin{table*}
\centering

\begin{tabular}{lllll}  
\hline
\textbf{Source URI} & \textbf{Target URI} & \textbf{Type of redirect} \\
\hline
http://blogs.nasa.gov/ & https://blogs.nasa.gov/ & HTTP to HTTPS Redirect \\
\hline
http://ecogeneration.com.au/ & https://www.ecogeneration.com.au/ & Non-WWW to WWW Redirect \\
\hline
https://www.langeley.edu.ar/ & https://langeley.edu.ar/ & WWW to Non-WWW Redirect \\
\hline
http://2a7t6cohz0-games.playsbo.com/ru-ru/ & http://2a7t6cohz0-games.playsbo.com/ru-RU/ & CASE Change Redirect \\
\hline
https://123moviesvf.com/ & http://123moviesvf.com/ & HTTPS to HTTP Redirect \\
\hline

\end{tabular}
    \caption{Examples of canonical redirects types}
    \label{tab:Ctypes}
\end{table*}

Non-canonicalized redirects occur when URIs are redirected to different URIs without establishing a canonical version. For example, If \texttt{\url{http://example.com/page1}} redirects to \texttt{\url{http://example.com/page2}} and \texttt{\url{http://example.com/page2}} further redirects to \texttt{\url{http://example.com/page3}}, but none of these URIs are identified as the canonical URI. Table~\ref{tab:NCtypes} shows various examples of non-canonical redirects.  Unlike canonical redirects, these redirections can introduce challenges in web archiving, as they may lead to instability in archived records. AlSum et al.~\cite{Ahmed2013redirection} found that nearly half of the studied URIs in archives were unstable, complicating the retrieval, and proposed policies to address these redirection challenges. Kelly et al. \cite{Mkelly:URIcanonicalization,Mkelly:URIcanonicalization_techreport} explored how URI canonicalization impacts the count of archival web captures in TimeMaps. Their research found that many redirects led to misleading counts, complicating the archiving process and the retrieval of accurate mementos. This underscores the importance of careful management and canonicalization of URIs to ensure the integrity of archived web resources.

\begin{table*}
\centering

\begin{tabular}{lllll}
\hline
\textbf{Source URI}                 & \textbf{Target URI}                      & \textbf{Type of non canonical redirect} \\ \hline
http://www.philosophie.lmu.de/      & https://www.philosophie.uni-muenchen.de/ & Domain Change Redirect                  \\ \hline
http://zoje-america.com/new/manufacturer/ & http://zoje-america.com/manufacturer/ & Old to New Page Redirect          \\ \hline
http://abackwardsstory.blogspot.cz/ & http://abackwardsstory.blogspot.com/     & TLD Change Redirect                     \\ \hline
https://radio.wosu.org/             & https://news.wosu.org/                   & Subdomain Change Redirect               \\ \hline
https://holymanga.net/              & https://w31.holymanga.net/               & Main domain to Subdomain Redirect       \\ \hline
http://research.louisville.edu/           & https://louisville.edu/research/      & Subdomain to Main Domain Redirect \\ \hline
\end{tabular}
    \caption{Examples of non-canonical redirects types}
    \label{tab:NCtypes}
\end{table*}

Previous studies have explored various dimensions of URI redirection. For example, Kline et al. \cite{Kline2019} conducted a comprehensive analysis of the World Wide Web's structure and dynamics by examining over 1 trillion URIs. Their study highlighted how URI traversal patterns differ significantly from hyperlink connectivity, underscoring the complex behaviors underlying web navigation. This complexity is further compounded by the prevalence of canonical and non-canonical redirects, which can obscure the true structure of web content. Another critical aspect of URI redirection is its impact on web performance and user experience. Lee et al.  \cite{Lee2009} introduced the concept of "soft errors" in redirections, where a URI redirects to a page that returns a 200 OK status but contains no relevant content. These soft errors degrade the performance of web search engines and can lead to a poor user experience. This issue is particularly relevant in the context of large-scale web crawling and SEO. We also analyzed a sample of our redirects for "soft errors." User web experiences are further influenced by factors such as regional censorship, where specific content is restricted based on location, or by personalized elements tied to user behavior, preferences, and history. Singh et al. \cite{singh2020} observed that censors often use HTTP redirection to block content, redirecting users to URLs displaying censorship notices. Additionally, the security implications of redirects are crucial to consider. A redirection chain, comprising a sequence of HTTP requests and responses, can be compromised if even a single link within the chain employs insecure HTTP protocols \cite{chang2017}. This vulnerability underscores the need for secure practices in designing and managing redirection mechanisms.

\section{Methodology}\label{sec:method}
For our previous research \cite{TrendMachine:2023,nypwfindingsblog2024}, we created a dataset of archived web pages by sampling IA’s Zipnum index file \cite{PywbIndexing}. This sample consists of TimeMaps for 27.3 million URIs first archived between 1996--2021, encompassing 3.4 billion mementos and 7.7 million unique hosts \cite{Samplingtechreport:2025}. In collaboration with IA, we crawled the dataset in June 2023 to determine the dead/alive status of each URI, but we did not follow redirects. 
We found that 11.7 million ($\sim$40\%) of the URIs were redirecting. When we re-crawled only the redirecting URIs in September 2023, we found that 11 million were still redirecting. The dataset is publicly available at the Internet Archive~\cite{crawldata_IA}. We removed approximately 1.5 million URIs that the crawler partially followed before encountering invalid redirects, resulting in a failure to produce a definitive terminating status code. Additionally, we removed 6,068 URIs that did not terminate after 10 hops. This resulted in a final dataset of 9.5 million unique redirecting URIs, representing over 2.6 million unique domains.

\subsection{Re-Crawling Redirecting URIs}

In September 2023, we performed a re-crawl of the 11.7 million redirecting URIs using IA's Heritrix. Unlike the June 2023 crawl, this crawl followed redirections up to 10 hops. We categorized each URI as success, redirect, or error based on the HTTP status code \cite{MDNHttpStatus} returned. URIs with HTTP status codes in the 2xx range were labeled as success and the 3xx range as redirect. Everything else was categorized as error, including client errors (4xx), server errors (5xx), DNS failures, HTTP connection errors, and any other error state encountered.
We found that 744,244 of the URIs were no longer redirecting, with 17\% terminating successfully and the remaining resulting in an error state. This left us with a dataset of 10,975,138 redirecting URIs for further analysis.

\subsection{Following Redirecting URIs}
\begin{figure*}[htbp]
  \centering
  \frame{\includegraphics[width=\linewidth]{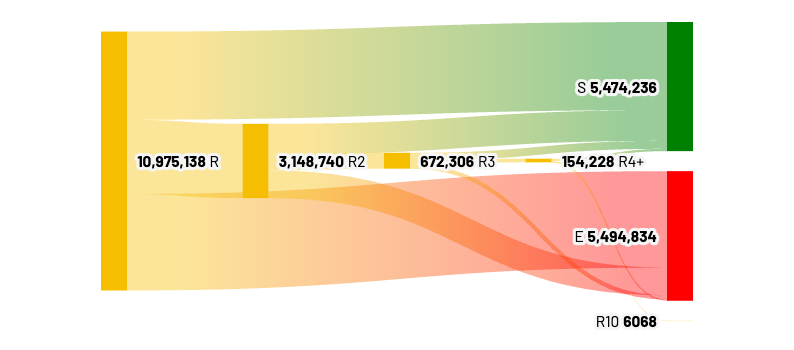}}
  \caption{Journey of 11 million URIs through multiple redirects, labeled as R (initial redirects) to R4+ (redirection chains comprising four or more stages). Each flow concludes in either success (S) or error (E) states, highlighting the outcomes of these chains. Of the 11 million initial redirects, half successfully reach their final destination, while the other half result in errors. The diagram also reveals details about the intermediate redirection stages: 3 million reach the second stage (R2), 672,306 advance to the third (R3). Notably, 6068 redirects, labeled as R10, were still redirecting at the tenth hop, indicating extended redirection paths that persist without resolution. }
  \Description{}
  \label{img:redirectjourney}
\end{figure*}

We analyzed the crawl logs to study the redirections of the remaining 11 million URIs, extracting the number of times each URI redirected and its final status. We calculated that out of 11 million URIs, 6.9 million URIs terminated after the first redirect, 2 million URIs terminated after the second redirect, 672,306 URIs redirected further. We observed that  1.5 million URIs were not followed by the crawler (these redirect errors will be explained further in Section \ref{sec:results}). 
Figure~\ref{img:redirectjourney} demonstrates the reduction in URIs with each redirect and the overall distribution of successful (green) and unsuccessful terminations/errors (red). The height of each colored section represents the URI count at each stage, showing how many URIs are redirected (up to four times). We observed that only 0.42\% of the redirects exceeded four hops without termination. This finding indicates that implementing a cap of five redirects during the crawl would sufficiently cover most cases.

Overall, we found that 9.5 million URIs terminated within 10 redirects, with 5.4 million terminating successfully. We found that 6,000 URIs encountered more than 10 consecutive redirects, leaving their termination status indeterminate. Additionally, 5.5 million URIs resulted in an error, including 1.5 million invalid redirects. We also used the Python package "tldextract" to extract the domain, host, and TLD of the source and target URIs \cite{TldExtract}. Our final dataset of 9.5 million unique redirecting URIs contained 2,637,166 unique domains.

\subsection{Distinguish Between Canonical and Non-Canonical Redirects}
We categorized our 9.5 million terminating URIs into canonicalized and non-canonicalized redirects. First, we converted both source and target URIs into their SURT form. If the SURT of the source and target URI was a exact match, we marked it as a canonicalized redirect. Otherwise, we marked it as a non-canonicalized redirect. We found that over half (6 million) of our 9.5 million redirecting URIs were classified as canonicalized, and 3.5 million URIs were classified as non-canonicalized redirects. 

\subsection{Crawl Results}\label{sec:results}

\begin{figure*}[htbp]
  \centering
  \frame{\includegraphics[width=0.6\linewidth]{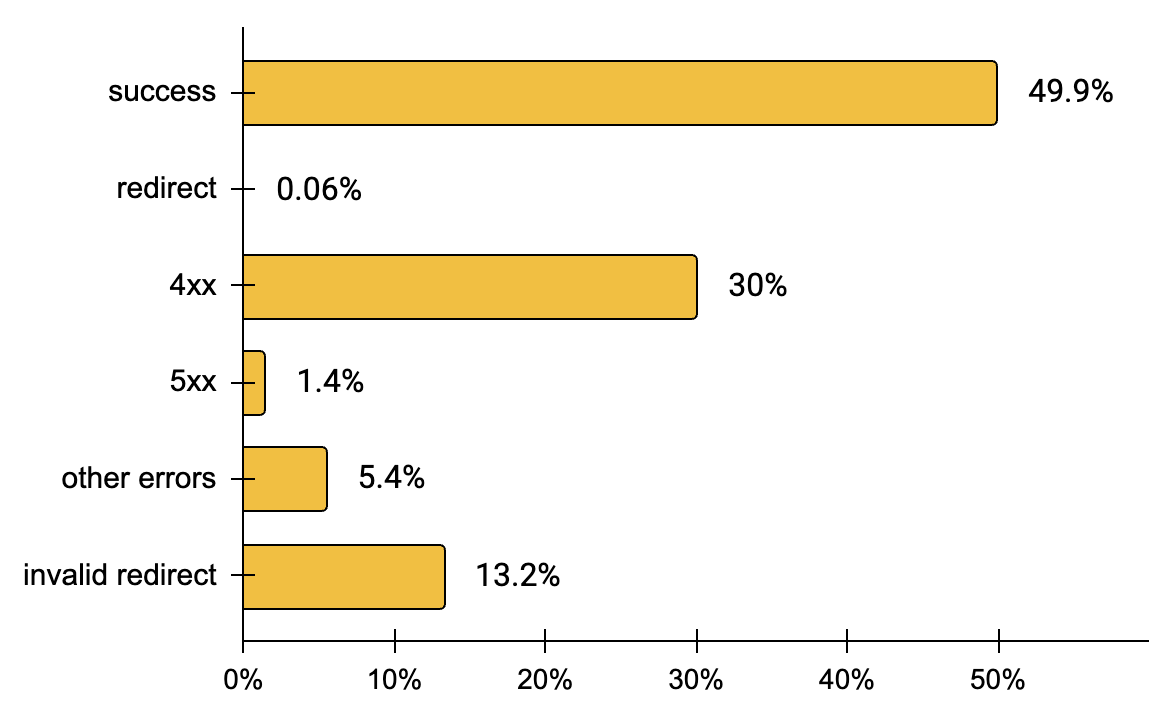}}
  \caption{Distribution of the terminated status codes of the 11 million redirecting URIs}
  \Description{}
  \label{img:terminatingStatus}
\end{figure*}

Figure~\ref{img:terminatingStatus} shows the distribution of the terminating status codes of the 9.5 million redirecting URIs and the 1.5 million unterminated URIs. We saw that 30\% of the URIs were redirected to a 4xx level status code, with a majority of those (86\%) being HTTP 404. We also observed that 665 URIs were redirected to 6xx-9xx status codes, which we grouped with DNS and HTTP connection errors as ``other error''. 

We found that the terminating status for 13.22\% of the 11 million URIs could not be determined as the crawler could not follow the redirect, which we termed as ``invalid redirect''. We examined a sample of these and found that the crawler stopped following them due to invalid HTTP Location headers (Figure~\ref{fig:invalidredirectexample_1}) or client-side redirect (Figure~\ref{fig:invalidredirectexample_2}), where a meta tag in the HTML document instructs the client to perform a redirection~\cite{w3c:metarefresh}. These redirects are not standard HTTP-level redirects and may introduce delays or parsing issues for the crawler. As a result, the crawler stops following these URIs.

\begin{figure*}
\begin{lstlisting}[numbers=none, backgroundcolor = \color{white}]
$ curl -iLs '010tarife.de/' 

HTTP/1.1 301 Moved Permanently
Date: Tue, 12 Sep 2023 17:24:53 GMT
Server: Apache/2.4.46 (Ubuntu)
<@\textcolor{red}{Location: https://www.{REQUEST\_URI} }@>      
Content-Length: 311
Content-Type: text/html; charset=iso-8859-1

\end{lstlisting}
\caption{Invalid redirect; a cURL request to a URI results in a 301 redirect with an invalid location header}
\Description{Invalid redirect; a cURL request to a URI results in a 301 redirect with an invalid location header}
\label{fig:invalidredirectexample_1}
\end{figure*}

\begin{figure*}
\begin{lstlisting}[numbers=none, backgroundcolor = \color{white}]
$ curl -iLs '013club.my163.com/' 

HTTP/1.1 301 Moved Permanently
Content-Type: text/html; charset=utf-8
<!DOCTYPE html>
<html>
<head>
  <title>Redirect 301</title>
  <@\textcolor{red}{<meta http-equiv="refresh" content="0;url=https://js.ninsud.com/download1/299\_0.html" />. }@>
</head>
<body>
</body>
</html>
\end{lstlisting}
\caption{Invalid redirect; a cURL request to a URI results in a 301 redirect without a location header, relying instead on an HTML meta-based redirection.}
\Description{Invalid redirect; a cURL request to a URI results in a 301 redirect without a location header, relying instead on an HTML meta-based redirection.}
\label{fig:invalidredirectexample_2}
\end{figure*}

\section{Analyzing Canonical Redirects}\label{sec:canon}

\begin{figure*}[htbp]
  \centering
  \frame{\includegraphics[width=0.7\linewidth]{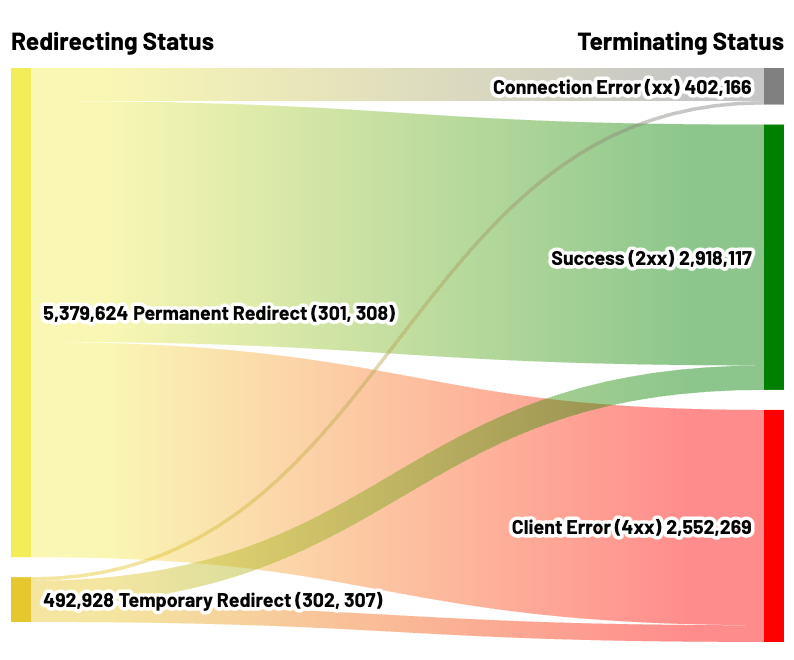}}
  \caption{
  The flow of canonical redirects from their initial redirect statuses to their final terminating statuses. The left side differentiates between two primary redirect categories: Permanent Redirects (301, 308), totaling 5.3 million occurrences, and Temporary Redirects (302, 307), totaling 492,928 occurrences. These flows then terminate on the right into three main status categories: Success (2xx) with 2.9 million occurrences, Client Errors (4xx) with 2.6 million occurrences, and Connection Errors (xx) with 402,166 occurrences.}
  \Description{
  The flow of canonical redirects from their initial redirect statuses to their final terminating statuses. The left side differentiates between two primary redirect categories: Permanent Redirects (301, 308), totaling 5.3 million occurrences, and Temporary Redirects (302, 307), totaling 492,928 occurrences. These flows then terminate on the right into three main status categories: Success (2xx) with 2.9 million occurrences, Client Errors (4xx) with 2.6 million occurrences, and Connection Errors (xx) with 402,166 occurrences.}

  \label{img:Canon_rolledup}
\end{figure*}

We analyzed the initial and terminating status codes of 6 million canonical redirects to understand their behavior and outcomes. Figure~\ref{img:Canon_rolledup} illustrates the flow of these redirects, showing their termination across various categories: successful requests (2xx), client errors (4xx), and non-standard errors ("xx"), which include issues such as DNS failures or connection errors.
The redirecting status codes in our dataset include 301 (Moved Permanently), 302 (Found), 303 (See Other), 307 (Temporary Redirect), and 308 (Permanent Redirect) \cite{MDN_HTTP_Redirections}. For simplicity, we grouped permanent redirects (301, 308) and temporary redirects (302, 307) together, as 308 and 307 were relatively infrequent. Additionally, we removed 303 redirects (9,617 occurrences) due to their low frequency compared to other redirect types.

Figure~\ref{img:Canon_rolledup} highlights that permanent redirects are more prevalent than temporary redirects in canonical redirection. This finding aligns with the intended purpose of canonical redirects, which aim to provide stable, long-term pathways for users and search engines to access the correct resource \cite{Redirection_Codes}. However, the presence of some temporary redirects in canonical redirection suggests that certain URIs are subject to conditional or temporary changes. While temporary redirects may serve specific purposes, their overuse in canonical contexts can have significant implications. For instance, temporary redirects signal to search engines that the change is not permanent, potentially delaying or preventing proper indexing of the target URI, as well as splitting PageRank calculations across the URI variations \cite{Moz_Canonicalization}. In canonical contexts, overusing temporary redirects can lead to ambiguity and potential SEO issues, highlighting the need for timely updates to permanent redirects. 

Our findings reveal that 48.70\% of canonical redirects successfully resolve to a 2xx status, demonstrating their general effectiveness in directing users to the correct resource. However, a significant portion terminates in client errors (4xx) or server errors (5xx), highlighting potential issues such as link rot, outdated URIs, improper permissions, or server instability. Although we initially anticipated that most canonical redirects would resolve to 200 OK statuses, our analysis revealed a more complex and error-prone landscape.

We analyzed the number of hops involved in canonical redirects to understand the complexity of their redirection paths. We hypothesized that most canonical redirects would involve a single hop, such as redirecting from \texttt{http://} to \texttt{https://}. The data supported this hypothesis, as 5.2 million canonical redirects were single hop. However, we also observed 701,265 canonical redirects involving two hops, 40,882 with three hops, and 1,765 with four hops. Below, we provide an example of a 4-hop canonical redirect ($S_0 \rightarrow R_1 \rightarrow R_2 \rightarrow R_3 \rightarrow R_4$) observed during our analysis:

\begin{itemize}
    \item S$_0$ (Source URI): \\ \texttt{http://148apps.com/app/305676364/hide}
    \item R$_1$: \texttt{\textbf{https}://148apps.com/app/305676364/hide} 
    \item R$_2$: \texttt{https://\textbf{www}.148apps.com/app/305676364/hide} 
    \item R$_3$: \texttt{\textbf{http}://www.148apps.com/app/305676364/hide\textbf{/}} 
    \item R$_4$ (Target URI): \\ \texttt{\textbf{https}://www.148apps.com/app/305676364/hide/} 
\end{itemize}
This sequence demonstrates a less efficient redirect chain where the original URI undergoes multiple transformations before arriving at the final canonical URI. Such cases can introduce delays, increase the likelihood of errors, and complicate the crawling process. 

\begin{figure*}[htbp]
  \centering
  \frame{\includegraphics[width=0.8\linewidth]{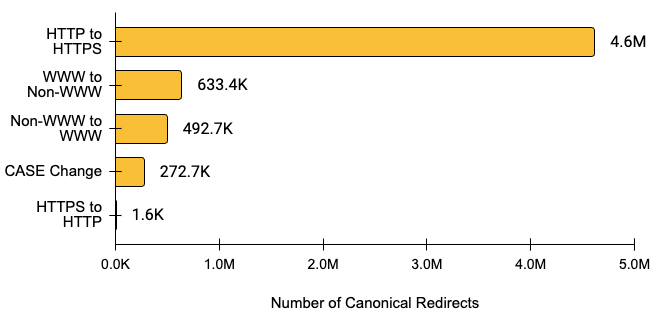}}
  \caption{Prevalence of different types of canonical redirects}
  \Description{Prevalence of different types of canonical redirects}
  \label{img:Canonicaltypes}
\end{figure*}

We analyzed the frequency of different types of canonicalized redirects in our sample. The analysis of 6 million canonical redirects strongly indicates that redirects from non-secure (HTTP) to secure (HTTPS) URIs are common, reflecting a broad adoption of secure protocols to enhance web security. The data reveals that HTTP to HTTPS redirects account for the majority, with approximately 4.6 million instances (Figure~\ref{img:Canonicaltypes}). Other types of redirects, such as those from WWW to non-WWW URIs and from non-WWW to WWW URIs, were far less common, with 633.4K and 492.7K instances, respectively. The rare occurrence of HTTPS to HTTP redirects (1.6K) further emphasizes the commitment to maintaining secure connections.

\section{Analyzing Non-Canonical Redirects} \label{sec:non_canon}

\begin{figure*}[htbp]
  \centering
  \frame{\includegraphics[width=0.7\linewidth]{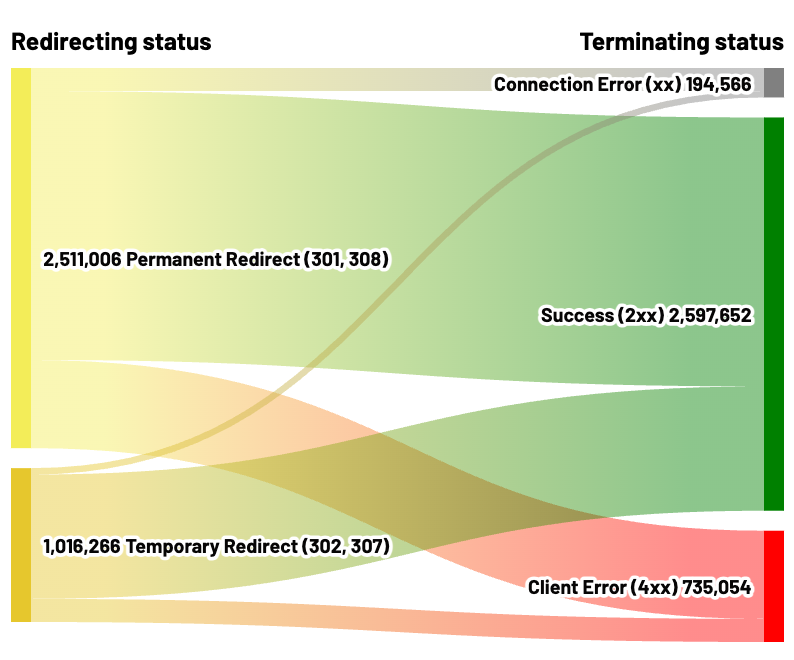}}
  \caption{
  The flow of non-canonical redirects from their initial redirecting statuses to their final terminating statuses. Two main redirect categories are shown: Permanent Redirects (301, 308) accounting for 2.5 million instances, and Temporary Redirects (302, 307) with 1 million instances. The terminating outcomes are classified into Success (2xx) with 2.5 million instances, Client Errors (4xx) with 735,054 instances, and Connection Errors (xx) with 194,566 instances. }
  \Description{The flow of non-canonical redirects from their initial redirecting statuses to their final terminating statuses. Two main redirect categories are shown: Permanent Redirects (301, 308) accounting for 2.5 million instances, and Temporary Redirects (302, 307) with 1 million instances. The terminating outcomes are classified into Success (2xx) with 2.5 million instances, Client Errors (4xx) with 735,054 instances, and Connection Errors (xx) with 194,566 instances.}
  \label{img:NonCanon_rolledup}
\end{figure*}

We examined the initial and terminating status codes of our 3.5 million non-canonical redirects. Figure~\ref{img:NonCanon_rolledup} offers insights into the behavior of non-canonical redirects. Around 73\% of these redirects successfully resolves to a 2xx status, indicating that users are generally reaching a destination. However, it is important to consider that even if non-canonical redirects terminate with a 200 OK status, it does not necessarily mean the original content has been moved to a new location. 
The redirect could lead to a root webpage instead of the original deep link, terminate in a parked page that offers no meaningful content, or may lead to a soft error page.
We explored popular target URIs in Section~\ref{sec:sinks} and discovered that although most terminated with a 200 OK status, the original content was effectively lost, and users are left with a misleading sense of successful navigation. This misrepresentation can mislead users and search engines, leading to issues in accurately indexing or preserving content.

When content moves across domains or to different servers, there is a risk that valuable information may be lost or become inaccessible over time, especially if these new destinations are not properly maintained or archived. The presence of network-related failures (xx errors) and client-side or server-side errors (4xx and 5xx) further highlights vulnerabilities that could hinder long-term content preservation.

\subsection{Impact of Redirection on URI Path Depth}

\begin{figure*}[htbp]
  \centering
\frame{\includegraphics[width=0.65\linewidth]{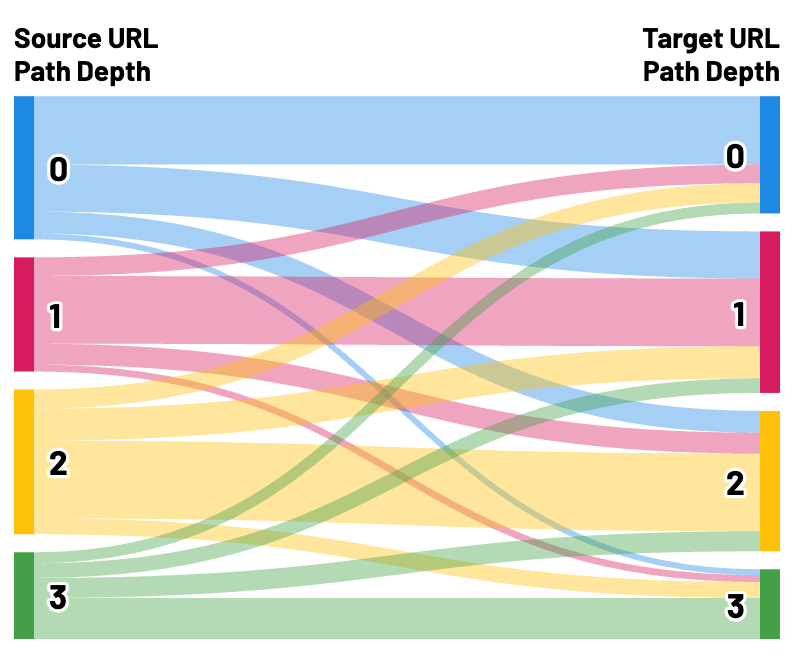}}
  \caption{
  The flow of URLs transitioning from source path depths to target path depths during non-canonical redirection. Each band represents a specific path depth level, numbered from 0 (root page) to 3, on both the source and target sides. The flow between left (source) and right (target) indicates changes in path depth as URLs are redirected.}  
  \Description{
  The flow of URLs transitioning from source path depths to target path depths during non-canonical redirection. Each band represents a specific path depth level, numbered from 0 (root page) to 3, on both the source and target sides. The flow between left (source) and right (target) indicates changes in path depth as URLs are redirected.}
  \label{img:SourceSinkpathdepth}
\end{figure*}

The analysis of redirection patterns across URI path depths provides valuable insights into the structural changes and practices involved in web redirection. Specifically, we examined the path depth of source URIs in comparison to their corresponding target URIs for non-canonicalized redirects. This approach reveals whether redirection practices typically maintain, simplify, or add complexity to the hierarchical structure of a URI. Figure~\ref{img:SourceSinkpathdepth} illustrates the changes in path depth between source and target URIs, ranging from 0 (root level) to 3 (deeply nested paths), highlighting how URL depths are altered during non-canonical redirection and providing insights into common URL restructuring and optimization practices in web navigation.

\subsubsection{Redirect to same path depth}
A key observation is the high proportion of redirects that maintain the same path depth. This consistency is seen in both root-level URIs and deeplink URIs, suggesting that many redirects aim to preserve the original hierarchical structure while possibly adjusting other aspects, such as domain names or language settings. For example, \texttt{\url{http://www.harrishometeam.com/}} redirects to \texttt{\url{https://timharris.kw.com/}}, keeping both URIs at the root level while rebranding the domain. Similarly, \texttt{\url{http://www.speaker-online.de/vifa-ase}} redirects to \texttt{\url{https://www.lautsprecherkauf.com/vifa-ase}}, preserving the deeper path structure but switching to a new domain.

\subsubsection{Redirect to smaller path depth}
Not all redirects, however, maintain the same depth. A significant number simplify their paths, moving from a deeplink to a root-level URI, as summarized in Figure~\ref{img:noncanon_pathdepth_summary}. These often involve legacy pages that are no longer relevant or whose content has been consolidated. For example, \texttt{\url{http://booked.jp/hotel/amerisuites-flagstaff-az-300216}} redirects to \texttt{\url{https://springhill-suites-flagstaff.booked.jp/}}, where an outdated hotel page now points to a central subdomain. Similarly, \texttt{\url{http://bokaa.com/info/17_1.htm}} redirects to \texttt{\url{https://www.bokaa.com/}}, effectively retiring a detailed page and redirecting traffic to their homepage. 

\subsubsection{Redirect to greater path depth}
Redirection to deeper paths also occur, indicating scenarios like directing users from root URIs to more targeted content or handling errors gracefully. For example, \texttt{\url{http://finylvinylrecords.com/}} redirects to \texttt{\url{https://www.hugedomains.com/domain_profile.cfm?d=finylvinylrecords.com}}, where a discontinued root-level page now points to a specific sales page for the domain. Similarly, \texttt{\url{http://www.hg.no/}} redirects to \texttt{\url{https://www.hg.no/403.shtml}}, a custom error page for an inaccessible domain. Another example is \texttt{\url{http://hiltonsuggests.hilton.com/}} redirecting to \texttt{\url{https://www.hilton.com/en/travel/}}, rerouting a defunct subdomain to a relevant deeplink within the main domain. These transitions often arise from restructuring efforts where the root-level page ceases to serve its original purpose and is redirected to a new platform, error page, or parked page. Another noteworthy example of redirects to deeper URIs involves URI shortening services, which often redirect from a concise, shallow URI to a much deeper, more complex destination. For instance, \texttt{\url{http://shorturl.at/aouS6}} redirects to \texttt{\url{https://www.figma.com/file/ZRT1lTxs8KQtlbvl33dMRb/alivio-landing-page-for-figma}}, preserving the target's detailed path structure while reducing the complexity of the URI presented to users. 

\begin{figure*}[htbp]
  \centering \frame{\includegraphics[width=0.65\linewidth]{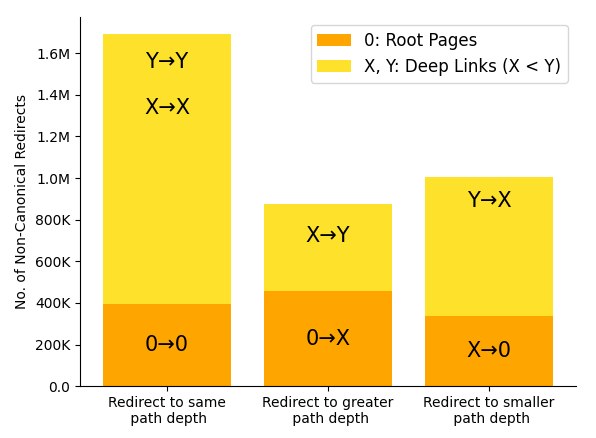}}
  \caption{Distribution of non-canonical redirects classified by changes in URI path depth. Redirects are divided into three groups: redirects maintaining the same path depth, redirects to a greater path depth, and redirects to a smaller path depth. Each group is further segmented into  redirects involving root pages, highlighted in orange (0→0, 0→X,X→0), and redirects involving deep links (X→X, X→Y, Y→X), where X and Y represent different depths with X < Y. }
    \Description{Image showing the distribution of non-canonical redirects classified by changes in URI path depth. Redirects are divided into three groups: redirects maintaining the same path depth, redirects to a greater path depth, and redirects to a smaller path depth. Each group is further segmented into  redirects involving root pages, highlighted in orange (0→0, 0→X,X→0), and redirects involving deep links (X→X, X→Y, Y→X), where X and Y represent different depths with X < Y. }
\label{img:noncanon_pathdepth_summary}
\end{figure*}

Out of the 3.5 million non-canonical redirects, 1.6 million maintain the same path depth, 400K of which involve root URIs, while 1.2 million are deeplinks. This preference for stability shows that webmasters value keeping existing structures intact, which helps preserve navigation and SEO value. At the same time, redirects to smaller depths simplify site structures by consolidating or retiring outdated content, while redirects to greater depths are typically employed to address specific needs, such as guiding users to targeted resources or managing discontinued pages. Figure~\ref{img:noncanon_pathdepth_summary} highlights these trends, showing a clear emphasis on maintaining or simplifying URIs rather than making them more complex. 

\subsection{Domain Changes}

\begin{figure*}
  \centering
\frame{\includegraphics[width=\linewidth]{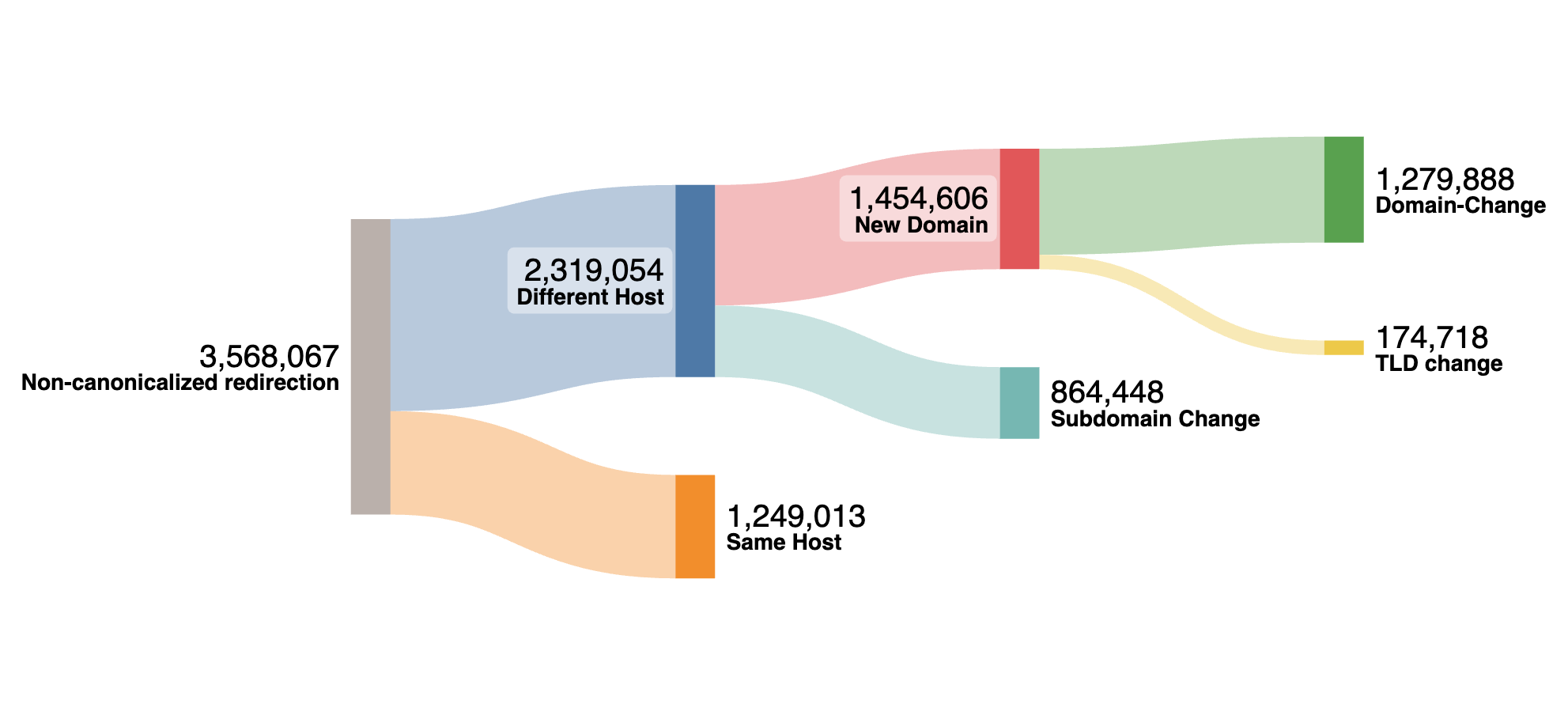}}
  \caption{Distribution of 3.5 million non-canonicalized redirections by host and domain changes. The majority (65\%) of redirections lead to a different host, with 37.3\% involving subdomain changes and 62.7\% transitioning to new domains.}  
  \Description{Image showing the distribution of 3.5 million non-canonicalized redirections by host and domain changes. The majority (65\%) of redirections lead to a different host, with 37.3\% involving subdomain changes and 62.7\% transitioning to new domains.}
\label{img:noncanonical_domain}
\end{figure*}

We investigated 3.5 million non-canonicalized redirects, categorizing them by the nature of the host and domain changes. Figure~\ref{img:noncanonical_domain}  visualizes the flow of these redirections, revealing that 65\% (2,319,054) redirect to a different host, while the remaining 35\% (1,249,013) occur within the same host. Same host redirections often occur to reroute users to updated page locations or to reflect website restructuring. For example, the URI \texttt{\url{https://boxhillindoorsports.com.au/sports-activities/bubble-soccer/book-birthday/}} now redirects to \texttt{\url{https://boxhillindoorsports.com.au/sports-and-activities/bubble-soccer/}}, a change likely aimed at consolidating related resources under a simplified structure. 

Among the different host redirects, 864,448 (37.3\%) involve a subdomain change, while 1,454,606 (62.7\%) redirect traffic to a new domain. Subdomain changes often occur when content is restructured or distributed across specialized subdomains. For instance, \texttt{\url{https://radio.wosu.org/}} now redirects to \texttt{\url{https://news.wosu.org/}}.
These changes preserve traffic flow within the same root domain but require proper implementation of canonical tags to prevent search engines from treating the subdomains as separate entities.

Of these new domain redirects, 1,279,888 point to entirely new domain names, while 174,718 involve changes only to the Top-Level Domain (TLD). The redirection of URIs to entirely new domain names demonstrates various strategic digital practices. 
For instance, URI shorteners serve as tools for simplifying complex web addresses, such as \texttt{\url{http://shorturl.at/afjCK}} redirection to an Instagram page \texttt{\url{https://www.instagram.com/memoriruangimajinasi/?igshid=MmU2YjMzNjRlOQ==}}. Additionally, 
corporate acquisitions can lead to organizations redirecting traffic to reflect changes in ownership, as seen in \texttt{\url{http://teragram.com/}} redirecting to \texttt{\url{https://www.sas.com/en_us/software/teragram.html}}. 
Finally, cases like \texttt{\url{http://advtise.net/}} redirecting to \texttt{\url{https://www.google.com/}} illustrate how expired or repurposed domains can redirect to search engines, maintaining some utility rather than leading to dead ends. 

TLD and domain changes may reflect rebranding efforts, market expansion, or strategies to optimize web visibility based on geographic or audience considerations \cite{Wu2011SearchEO}. For instance, \texttt{\url{http://irishpost.co.uk/}} now redirects to \texttt{\url{http://irishpost.com/}} as part of a rebranding effort, while \texttt{\url{http://www.adobe.co.in/}} redirects to \texttt{\url{https://www.adobe.com/in/}} to align with Adobe’s global domain structure while retaining localized content for Indian users. 

These patterns reflect a range of motivations and consequences. Organizations often use cross-domain redirects to consolidate traffic, implement rebranding efforts, or increase domain authority. When managed properly, these practices can enhance SEO performance. However, excessive or poorly managed redirects can dilute link equity, reducing search engine ranking effectiveness \cite{Sharma2022}. For example, if a website redirects users from pageA.com to pageB.com and then to pageC.com, search engines may fail to consolidate the link authority, leading to ranking losses.

The findings also highlight potential security risks. Redirects to new domains can obscure malicious activities, such as phishing or malware distribution, particularly when end-users are unaware of redirection intentions \cite{Wang2005}. For example, a user who redirects from \texttt{secure-login.com} to a fraudulent \texttt{secure-login.net} may unknowingly enter sensitive information into a malicious site. The complexity of redirection chains and lack of transparency may provide a cover for unethical practices, emphasizing the need for robust monitoring and ethical standards in web redirection management \cite{Zhang2017}.

\subsection{Analyzing Top Redirection Sink URIs}\label{sec:sinks}
We noticed that some of our source URIs terminated at a common target URI, which we refer to as a ``sink''. To understand this redirection pattern, we analyzed the most common sink URIs in our non-canonical redirects. Table~\ref{tab:topsinks_url} shows the top 15 sinks with the highest source URI frequency redirecting to these in non-canonical redirects sample.

\subsubsection{Consolidation of Organizational Domains}

We encountered instances where URIs from a specific organization or domain were redirected to a single sink. For example, numerous regional domains and subdomains associated with Allrecipes, such as \texttt{allrecipes.cn}, \texttt{allrecipes.asia}, and \texttt{allrecipes.co.in},  were being redirected to the primary global domain, \texttt{allrecipes.com}. This widespread redirection indicates a strategic consolidation of regional websites into a single, unified platform.

We saw another example of traffic consolidation in our top sinks but with a different purpose and context. We found various news and content-related domains, such as \texttt{palmbeachpost.com} and  \texttt{cjonline.com}, redirecting to the homepage of \texttt{\url{https://www.usatoday.com/}}. This pattern suggests that these domains, likely belonging to the same media network or ownership group, are directing their online presence toward USA Today’s main site. However, the fact that these URIs redirect to the homepage rather than specific content pages suggests that the original articles or resources may no longer be available. 
This broad redirection to the homepage can also indicate an effort to retain traffic from older or deprecated URIs by ensuring that users still reach the primary site, even if the content they seek is no longer accessible.

\begin{table*}[ht]
\begin{tabular}{|r|l|}
\hline
\textbf{Source URIs} & \textbf{Top Sinks}                                      \\
\hline
1242 & https://pharm-discount.net/?aff=1023/ \\ \hline
738 & https://www.000webhost.com/migrate?static=true \\ \hline
589 & https://www.allrecipes.com/ \\ \hline
583 & http{[}s{]}://www.google.com/ \\ \hline
554 & http://dfltweb1.onamae.com/ \\ \hline
489 & https://www.youtube.com/watch?v=oHg5SJYRHA0 \\ \hline
476 & https://archive.org/about/404.html \\ \hline
365 & https://www.usatoday.com/ \\ \hline
338 & https://orghost.ru/ \\ \hline
333 & https://twitter.com/login \\ \hline
327 & https://mercadolibre.com/ \\ \hline
235 & https://w1.buysub.com/pubs/HR/A14/Hearst\_Subscription\_LP.jsp?cds\_mag\_code=A14\&cds\_page\_id=257255 \\ \hline
234 & https://www.wp.pl/?404\&src01=99f53 \\ \hline
232 & https://err.freewebhostingarea.com/404.html \\ \hline
221 & https://6789000000.com/register?id=19364165 \\ \hline
\end{tabular}
\caption{Top 15 most frequent sink URIs by source URI count}
\label{tab:topsinks_url}
\end{table*}

\subsubsection{Affiliate Marketing Sinks}
We also analyzed the URIs where multiple domains were redirecting to a single sink. Table~\ref{tab:topsinks_domain} presents the top 15 sinks in our sample, ranked by the frequency of source domains redirecting to them. Our topmost sink is the \texttt{\url{https://pharm-discount.net/?aff=1023/}}, an online pharmacy store, to which 1,242 unique URIs from 323 different domains were redirected (it is the top sink in Table \ref{tab:topsinks_url} as well). We observed that all the source domains directing traffic to this sink were also related to pharmacy websites, such as \texttt{24h-canadian-pharmacy.com}, \texttt{grandhealthstore.com}, and \texttt{fastpills-online.com}. This online pharmacy store consolidates traffic from multiple sources, likely representing various smaller or niche online pharmacies, into one main platform that offers discounted medicines. This approach is often seen in affiliate marketing, where multiple websites drive traffic to a single commercial site, enhancing visibility and potential customer base. The parameter \texttt{?aff=1023} in the URL likely serves as a tracking identifier, commonly used in affiliate marketing and user tracking systems \cite{koop2020}. 

\begin{table*}
\begin{tabular}{|r|p{0.75\linewidth}|}
\hline
\textbf{Source Domains} & \textbf{Top Sinks} \\
\hline
323 & https://pharm-discount.net/?aff=1023/ \\ \hline
282 & https://twitter.com/login \\ \hline
273 & http{[}s{]}://www.google.com/ \\ \hline
141 & http://dfltweb1.onamae.com/ \\ \hline
131 & https://archive.org/about/404.html \\ \hline
86 & https://6789000000.com/register?id=19364165 \\ \hline
82 & https://www.youtube.com/watch?v=oHg5SJYRHA0 \\ \hline
79 & http://affordablewebhosting.com/adscheaper.htm \\ \hline
77 & http://127.0.0.1/ \\ \hline
54 & https://www.ovhcloud.com/en-gb/mail/ \\ \hline
53 & http://errdoc.gabia.io/404.html \\ \hline
51 & https://www.yahoo.com/?spiders \\ \hline
51 & http://www.bing.com/ \\ \hline
48 & https://www.usatoday.com/ \\ \hline
40 & https://www.000webhost.com/migrate?static=true \\ \hline
\end{tabular}
\caption{Top 15 most frequent sink URIs by source domain count.}
\label{tab:topsinks_domain}
\end{table*}

\subsubsection{Social Media and Login Redirects}
Our top sinks include login pages, such as \texttt{\url{https://twitter.com/login}}. This is primarily due to ``share'' buttons embedded in many articles, which direct users to post content on social media platforms like Twitter. When a web archive crawler encounters these embedded share links, the social media's web server typically issues a redirect to the login page because the crawler, operating without logging in, cannot authenticate the request. Consequently, crawlers navigating these embedded share links are consistently redirected to the login interface, capturing the login page in the archive instead of the intended content \cite{bragg-jcdl23}.

\subsubsection{Inactive or Expired Domain Sinks}
We also encountered hosting or domain registration services such as \texttt{\url{http://dfltweb1.onamae.com/}}, \texttt{\url{https://www.000webhost.com/migrate?static=true}}, and \texttt{\url{http://affordablewebhosting.com/adscheaper.htm}} in our sinks. These sinks redirect traffic from inactive, expired, or deactivated domains to a generic page provided by the hosting or domain registration service. These sinks often occur when websites hosted on these platforms are discontinued, deactivated, or not renewed by their owners. We also found that the sink \texttt{\url{https://www.000webhost.com/migrate?static=true}} now returns a 403 Forbidden error, indicating that access to this page is restricted. As a result, URIs redirecting to this sink lead users to a non-functional page, effectively breaking the redirects and diminishing their utility. This suggests that the original purpose of the migration page has ended, or it has been intentionally disabled.

\subsubsection{Loopback Sink}
Other interesting sink we observed was \texttt{\url{http://127.0.0.1/}}. Redirecting URIs to the localhost is a method used to disable broken, or inactive links by pointing users to their own computer’s loopback address. While this approach prevents users from accessing unsafe or non-existent pages, it creates issues for web crawlers. Crawlers cannot access the intended content and may interpret the redirect as an error, negatively impacting the website’s search rankings and visibility. It can also indicate an attack on the local (crawling) machine \cite{cushman:2017, Leitschuh:2019}. A better alternative would be to redirect to an informative error page with appropriate HTTP status codes (such as \texttt{404 Not Found} or \texttt{410 Gone}) and to maintain an updated sitemap, ensuring clarity for both users and crawlers without causing unnecessary confusion.

\subsubsection{Prank and Meme Sink}
\begin{figure}[htbp]
  \centering
\frame{\includegraphics[width=\linewidth]{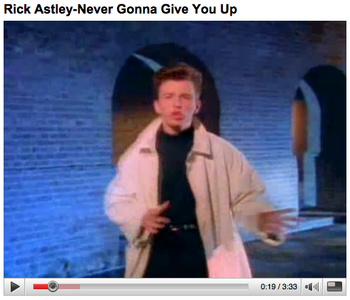}}
  \caption{A screenshot from the music video of "Never Gonna Give You Up."}
  \Description{A screenshot from the music video of "Never Gonna Give You Up."}
  \label{img:RickRoll}
\end{figure}

One sink that we found in the top five in terms of source URIs (489) and top 10 in terms of source domains (82) was notable -- \texttt{\url{https://www.youtube.com/watch?v=oHg5SJYRHA0}}, a video with almost 100 million views \cite{rickrolling_blog}. The sink is famously associated with the internet phenomenon "Rickrolling." When users click on a link expecting to be directed to a specific content, they are instead redirected to this YouTube video, which features the music video for Rick Astley's 1987 hit song "Never Gonna Give You Up" (Figure~\ref{img:RickRoll}). In this case, all source URIs redirecting to this video were associated with adult content websites, suggesting a deliberate effort to divert users from explicit material to an unrelated, benign video. This type of redirection is typically done as a prank, taking advantage of the unexpected nature of the redirect to surprise the user. 

\subsubsection{Search Engine Sinks}

We witnessed search engine web pages in our top sinks such as \texttt{\url{https://www.yahoo.com/?spiders}}, \texttt{\url{http://www.bing.com/}}, and \texttt{\url{http[s]://www.google.com/}}. The Yahoo sink, marked by the \texttt{?spiders} query does not alter the content displayed and functions as a standard Yahoo homepage. This query parameter is likely to be used for tracking or analytics purposes. The search engine web page sinks function as general-purpose landing pages for a variety of redirected URIs, ensuring that both human users and automated traffic are directed to the main homepage. The redirection of URIs from various sources to the root page of the search engine web page suggests that these original URIs are either outdated, broken, or no longer maintained. This pattern is common when the original content or websites have been taken down, and the domain owner has opted to redirect traffic to a general search engine.

 Google often redirects old or discontinued services, such as Google+, to their main search page once they are no longer active. For instance, we observed that various Google subdomains, like \texttt{clients1.google.com.bz} and \texttt{desktop.google.ca}, as well as regional domains like \texttt{google.ca} and \texttt{google.co.uk}, are redirected to the main homepage, \texttt{\url{http://www.google.com/}}. 
 This suggests that these specific subdomains or localized services are no longer active and have been deprecated in favor of centralizing traffic to a single, global domain. 
 
 We noticed that \texttt{http://www.google.com/} did not redirect to \texttt{http\textbf{s}://www.google.com/} , creating two separate sinks, which we confirmed using \texttt{curl}. We combined their source URI frequencies under \texttt{http[s]://www.google.com/} in Table~\ref{tab:topsinks_url} and Table~\ref{tab:topsinks_domain}, as both sinks served the same purpose. Interestingly, while \texttt{curl} revealed no redirection for \texttt{http://www.google.com/}, testing in an interactive browser showed that \texttt{http://www.google.com/} did indeed redirect to \texttt{http\textbf{s}://www.google.com/}. This discrepancy highlights a key difference between how crawlers or automated tools perceive the web compared to human users, who interact with browsers that handle redirects dynamically. Interactive browsers may trigger more redirects due to the execution of JavaScript or other browser-based mechanisms, meaning that the number of redirects observed using our crawler could serve as a conservative floor value.

\subsubsection{Custom Error Page Sinks}\label{subsec:error_sinks}

Tables~\ref{tab:topsinks_url} and ~\ref{tab:topsinks_domain} contains sinks that contain the characters ``404'' in their URIs, such as \texttt{\url{https://archive.org/about/404.html}},
\texttt{\url{http://errdoc.gabia.io/404.html}},
\texttt{\url{https://www.wp.pl/?404\&src01=99f53}},
 and \texttt{\url{https://err.freewebhostingarea.com/404.html}}. These sinks are custom 404 web pages, where users and crawlers land when attempting to access non-existent or broken links.  

The source URIs redirecting to \texttt{\url{https://archive.org/about/404.html}} share a common characteristic: they all use the \texttt{.work} TLD. The \texttt{.work} TLD, like other newer domain extensions, is often used for temporary or experimental  websites because of its affordability. The prevalence of \texttt{.work} domains among these redirects suggests that these sites may have been created for short-term projects, spam, or low-quality content, and are now either abandoned or improperly managed. Consequently, the broken or non-existent pages from these \texttt{.work} domains lead users and web crawlers to the IA's 404 error page.

The redirects to \texttt{\url{http://errdoc.gabia.io/404.html}} originate from a variety of South Korean websites, many of which also appear to be outdated or no longer active. These sites, hosted by Gabia, a South Korean web hosting service, redirect to Gabia’s generic 404 page when encountering missing pages.  Alkwai et al. \cite{lulwah2017} also identified Gabia’s custom error page \texttt{\url{http://errdoc.gabia.io/403.html}} among the top 10 most archived Korean URIs. Their study revealed that five of the top 10 most archived Korean URIs were custom error pages. When the crawler attempted to archive these outdated or missing source URIs, it was redirected to error pages, resulting in the repeated archiving of these pages. This not only wastes storage resources but also undermines archival efforts to combat link rot by preserving invalid URIs as though they were legitimate pages.

Similarly, the source URIs redirecting to \texttt{\url{https://www.wp.pl/?404\&src01=99f53}} predominantly come from websites once associated with WP.pl, short for Wirtualna Polska, a major Polish web portal. These includes  platforms like Pinger.pl (a personal blogging service) and Webpark.pl (a website hosting service), along with other sites that now seem inactive or outdated \cite{reddit_pingerpl}. Consequently, when users or crawlers attempt to access these URIs, they are redirected to wp.pl's generic 404 error page.

The source URIs that redirect to \texttt{\url{https://err.freewebhostingarea.com/404.html}} are predominantly hosted on free web hosting services, including domains such as 6te.net, freetzi.com, and orgfree.com. These websites, which likely served as personal projects, small forums, or niche content sites, now appear inactive or abandoned. As a result, requests to these URIs are redirected to Free Web Hosting Area's default 404 error page. This phenomenon indicates a broader issue in free hosting environments, where the lack of ongoing maintenance leads to many dead links.

Overall, analyzing these error sinks highlights the common issue of link rot across various domains and hosting environments. Whether due to the transient nature of content on low-cost hosting platforms, or the challenges of maintaining large web portals, the frequent redirection to custom 404 error pages reflects the broader problem of rotting web content over time. Furthermore, many of these error sinks, indicating missing or non-existent content, are soft errors or soft 404s, i.e., they return an HTTP 200 OK status code instead of the expected 404 Not Found. We analyzed the custom 404 URIs in our sample to determine how many are classified as soft 404s. Tables~\ref{tab:errorsinks} demonstrates the most frequent custom error pages in our dataset. Among the target URIs in our dataset, we identified 62,000 custom 404 URIs. Upon examining the terminating status of these URIs, we found that only 46\% returned a 404 Not Found status, while 47\% were incorrectly terminated with a 200 OK status despite being 404 error pages.

\begin{table*}[]
\begin{tabular}{|c|l|c|}
\hline
\textbf{Source URLs} & \multicolumn{1}{c|}{\textbf{Custom Error SINK}}               & \textbf{Status Code} \\ \hline
476 & https://archive.org/about/404.html          & 200 \\ \hline
234 & https://www.wp.pl/?404\&src01=99f53         & 200 \\ \hline
232 & https://err.freewebhostingarea.com/404.html & 200 \\ \hline
154                  & http://chaturbate.com/affiliates/in/grq0/KZiNo/?track=404exit & 429                  \\ \hline
121 & http://stat.dyna.ultraweb.hu/404.php        & 404 \\ \hline
106 & https://openstax.org/general/cnx-404/       & 200 \\ \hline
104 & https://www.premiumgfs.com/404/             & 200 \\ \hline
94  & https://404.onet.pl/                        & 404 \\ \hline
86  & https://blogs.miarroba.com/error\_404.php   & 404 \\ \hline
85  & http://srv4.sws.im/error/?c=404             & 200 \\ \hline
\end{tabular}
\caption{Top 10 custom error sink URIs by source URI count, along with their status codes showing how some of these custom error pages are soft 404s.}
\label{tab:errorsinks}
\end{table*}

\section{Future Work}\label{sec:fw}

This study highlights several promising directions for future research. A detailed analysis of target pages of the successful non-canonical redirects could clarify whether content has been relocated, serves as a placeholder, or represents a soft error page, deepening our understanding of link rot and redirection strategies. Additionally, examining the relationship between non-canonicalized redirects and archiving behavior could uncover patterns that inform more effective archival practices. This includes improving the integration of TimeMaps to enhance the retrieval of mementos when the original resource's URI has changed due to HTTP redirection. Analyzing longitudinal data on redirect patterns could provide deeper insights into how web resources evolve over time. Investigating the lifespan of different types of redirects, the emergence of sink URLs, and the frequency with which URIs change would help reveal broader trends in web maintenance, SEO strategies, and content persistence. Furthermore, the redirects dataset could be leveraged to identify patterns associated with malicious redirects, enabling the detection of misuse in phishing attacks and other security threats. A deeper exploration of these behaviors could support the development of more effective detection and mitigation strategies.

\section{Conclusions}\label{sec:conclusion}
Our study analyzed 11 million unique redirecting URIs to uncover key patterns in their usage and implications. About 50\% of these URIs successfully led to live webpages, while the remainder resulted in errors. A small fraction (0.06\%) exceeded 10 hops, and only 0.42\% surpassed four steps without termination, supporting the commonly used five-hop limit to prevent latency and manage crawl budgets effectively Additionally, 13.22\% of URIs had indeterminate termination statuses due to invalid redirects, highlighting technical inefficiencies that warrant attention.

We categorized redirects as either canonical or non-canonical. Canonical redirects, often aligned with SEO best practices, included 4.6 million HTTP-to-HTTPS transitions, signaling a shift toward secure web traffic. While canonical redirects were generally successful (48.70\% resolved to 2xx status codes), many resulted in client or server errors, reflecting link rot and server instability. Non-canonical redirects, which frequently involved domain changes (65\%), presented greater challenges, including frequent content loss, soft 404 errors, and inefficient traffic consolidation. Redirects to new domains also posed security risks, as they could obscure malicious activity and erode user trust. Redirect chains generally maintained or simplified path depth, reflecting efforts to consolidate content while avoiding unnecessary complexity.

A notable contribution of this study is the identification of redirection sink URIs. These sinks included login pages, custom error pages, and parked domains, serving varied roles such as traffic consolidation, affiliate marketing, and even internet pranks. However, their prevalence raises concerns about wasted crawler resources, inefficient traffic routing, and content decay. We also examined the widespread occurrence of soft 404 errors, finding that 47\% of custom 404 pages functioned as soft 404s. Such errors obscure the true status of redirected content and complicate digital preservation.

Our research identified strong indicators for detecting content drift or decay by analyzing source and target URIs. Redirects to root pages often signal content loss, while redirects from root pages to deeper links frequently indicate parked domains or custom error pages. Similarly, redirects leading to sink pages strongly suggest content loss. Target URIs resembling custom error pages but returning a 200 OK status code further serve as markers of soft 404 errors. These insights provide a practical framework for assessing the integrity of redirection chains without extensive content analysis.

This work offers valuable implications for webmasters, researchers, and digital archivists. Webmasters can optimize redirection strategies by minimizing multi-hop chains, addressing soft 404 errors, and managing domain transitions transparently. Researchers can build on these findings to deepen the understanding of web redirection and its impact on the online ecosystem. Digital archivists can apply these insights to improve content preservation amid increasingly complex redirection practices. By adopting efficient redirection practices and resolving technical inefficiencies, stakeholders can enhance web usability, optimize resources, and safeguard valuable content.

\begin{acks}
This work is supported in part by Protocol Labs and the Filecoin Foundation. Jake LaFountain from the Internet Archive provided valuable assistance with crawling redirecting URLs.
\end{acks}

\bibliographystyle{ACM-Reference-Format}
\bibliography{ref}


\begin{thebibliography}{50}


\ifx \showCODEN    \undefined \def \showCODEN     #1{\unskip}     \fi
\ifx \showDOI      \undefined \def \showDOI       #1{#1}\fi
\ifx \showISBNx    \undefined \def \showISBNx     #1{\unskip}     \fi
\ifx \showISBNxiii \undefined \def \showISBNxiii  #1{\unskip}     \fi
\ifx \showISSN     \undefined \def \showISSN      #1{\unskip}     \fi
\ifx \showLCCN     \undefined \def \showLCCN      #1{\unskip}     \fi
\ifx \shownote     \undefined \def \shownote      #1{#1}          \fi
\ifx \showarticletitle \undefined \def \showarticletitle #1{#1}   \fi
\ifx \showURL      \undefined \def \showURL       {\relax}        \fi
\providecommand\bibfield[2]{#2}
\providecommand\bibinfo[2]{#2}
\providecommand\natexlab[1]{#1}
\providecommand\showeprint[2][]{arXiv:#2}

\bibitem[Agata et~al\mbox{.}(2014)]%
        {agata2014lifespan}
\bibfield{author}{\bibinfo{person}{Teru Agata}, \bibinfo{person}{Yosuke Miyata}, \bibinfo{person}{Emi Ishita}, \bibinfo{person}{Atsushi Ikeuchi}, {and} \bibinfo{person}{Shuichi Ueda}.} \bibinfo{year}{2014}\natexlab{}.
\newblock \showarticletitle{Life Span of Web Pages: A Survey of 10 Million Pages Collected in 2001}. In \bibinfo{booktitle}{\emph{Proceedings of the 14th ACM/IEEE-CS Joint Conference on Digital Libraries (JCDL 2014)}}. \bibinfo{publisher}{IEEE}, \bibinfo{address}{London, UK}, \bibinfo{pages}{463--464}.
\newblock
\urldef\tempurl%
\url{https://doi.org/10.1109/JCDL.2014.6970226}
\showDOI{\tempurl}


\bibitem[Alam et~al\mbox{.}(2023)]%
        {TrendMachine:2023}
\bibfield{author}{\bibinfo{person}{Sawood Alam}, \bibinfo{person}{Kritika Garg}, \bibinfo{person}{Michele~C. Weigle}, \bibinfo{person}{Michael~L. Nelson}, \bibinfo{person}{Mark Graham}, {and} \bibinfo{person}{Dietrich Ayala}.} \bibinfo{year}{2023}\natexlab{}.
\newblock \showarticletitle{TrendMachine: A Temporal Webpage Resilience Portal}. In \bibinfo{booktitle}{\emph{Proceedings of the ACM/IEEE Joint Conference on Digital Libraries (JCDL)}}. \bibinfo{publisher}{IEEE Press}, \bibinfo{address}{Santa Fe, New Mexico, USA}, \bibinfo{pages}{93--97}.
\newblock
\urldef\tempurl%
\url{https://doi.org/10.1109/JCDL57899.2023.00023}
\showDOI{\tempurl}


\bibitem[Alkwai et~al\mbox{.}(2017)]%
        {lulwah2017}
\bibfield{author}{\bibinfo{person}{Lulwah~M. Alkwai}, \bibinfo{person}{Michael~L. Nelson}, {and} \bibinfo{person}{Michele~C. Weigle}.} \bibinfo{year}{2017}\natexlab{}.
\newblock \showarticletitle{Comparing the Archival Rate of Arabic, English, Danish, and Korean Language Web Pages}.
\newblock \bibinfo{journal}{\emph{ACM Transactions on Information Systems}} \bibinfo{volume}{36}, \bibinfo{number}{1}, Article \bibinfo{articleno}{1} (\bibinfo{date}{June} \bibinfo{year}{2017}), \bibinfo{numpages}{34}~pages.
\newblock
\showISSN{1046-8188}
\urldef\tempurl%
\url{https://doi.org/10.1145/3041656}
\showDOI{\tempurl}


\bibitem[AlSum et~al\mbox{.}(2013)]%
        {Ahmed2013redirection}
\bibfield{author}{\bibinfo{person}{Ahmed AlSum}, \bibinfo{person}{Michael~L. Nelson}, \bibinfo{person}{Robert Sanderson}, {and} \bibinfo{person}{Herbert Van~de Sompel}.} \bibinfo{year}{2013}\natexlab{}.
\newblock \showarticletitle{Archival HTTP redirection retrieval policies}. In \bibinfo{booktitle}{\emph{Proceedings of the 22nd International Conference on World Wide Web}} \emph{(\bibinfo{series}{WWW '13 Companion})}. \bibinfo{publisher}{Association for Computing Machinery}, \bibinfo{address}{Rio de Janeiro, Brazil}, \bibinfo{pages}{1051–1058}.
\newblock
\showISBNx{9781450320382}
\urldef\tempurl%
\url{https://doi.org/10.1145/2487788.2488117}
\showDOI{\tempurl}


\bibitem[Archive(2024)]%
        {Heritrix3Wiki}
\bibfield{author}{\bibinfo{person}{Internet Archive}.} \bibinfo{year}{2024}\natexlab{}.
\newblock \bibinfo{title}{Heritrix 3 Wiki}.
\newblock \bibinfo{howpublished}{GitHub Wiki}.
\newblock
\urldef\tempurl%
\url{https://github.com/internetarchive/heritrix3/wiki}
\showURL{%
\tempurl}


\bibitem[Bar-Yossef et~al\mbox{.}(2004)]%
        {bar2004sic}
\bibfield{author}{\bibinfo{person}{Ziv Bar-Yossef}, \bibinfo{person}{Andrei~Z. Broder}, \bibinfo{person}{Ravi Kumar}, {and} \bibinfo{person}{Andrew Tomkins}.} \bibinfo{year}{2004}\natexlab{}.
\newblock \showarticletitle{Sic transit gloria telae: towards an understanding of the web's decay}. In \bibinfo{booktitle}{\emph{Proceedings of the 13th International Conference on World Wide Web}} (New York, NY, USA) \emph{(\bibinfo{series}{WWW '04})}. \bibinfo{publisher}{Association for Computing Machinery}, \bibinfo{address}{New York, NY, USA}, \bibinfo{pages}{328–337}.
\newblock
\showISBNx{158113844X}
\urldef\tempurl%
\url{https://doi.org/10.1145/988672.988716}
\showDOI{\tempurl}


\bibitem[Berners-Lee et~al\mbox{.}(1994)]%
        {BernersLee1994}
\bibfield{author}{\bibinfo{person}{Tim Berners-Lee}, \bibinfo{person}{Robert Cailliau}, \bibinfo{person}{Ari Luotonen}, \bibinfo{person}{Henrik~Frystyk Nielsen}, {and} \bibinfo{person}{Arthur Secret}.} \bibinfo{year}{1994}\natexlab{}.
\newblock \showarticletitle{The World-Wide Web}.
\newblock \bibinfo{journal}{\emph{Commun. ACM}} \bibinfo{volume}{37}, \bibinfo{number}{8} (\bibinfo{date}{August} \bibinfo{year}{1994}), \bibinfo{pages}{76–82}.
\newblock
\showISSN{0001-0782}
\urldef\tempurl%
\url{https://doi.org/10.1145/179606.179671}
\showDOI{\tempurl}


\bibitem[Berners-Lee et~al\mbox{.}(2005)]%
        {rfc3986}
\bibfield{author}{\bibinfo{person}{Tim Berners-Lee}, \bibinfo{person}{Roy~T. Fielding}, {and} \bibinfo{person}{Lawrence Masinter}.} \bibinfo{year}{2005}\natexlab{}.
\newblock \bibinfo{title}{{Uniform Resource Identifier (URI): Generic Syntax, RFC 3986}}.
\newblock
\newblock
\urldef\tempurl%
\url{https://www.rfc-editor.org/rfc/rfc3986}
\showURL{%
\tempurl}


\bibitem[Bragg et~al\mbox{.}(2023)]%
        {bragg-jcdl23}
\bibfield{author}{\bibinfo{person}{Haley Bragg}, \bibinfo{person}{Himarsha Jayanetti}, \bibinfo{person}{Michael~L. Nelson}, {and} \bibinfo{person}{Michele~C. Weigle}.} \bibinfo{year}{2023}\natexlab{}.
\newblock \showarticletitle{Less than 4\% of Archived Instagram Account Pages for the Disinformation Dozen are Replayable}. In \bibinfo{booktitle}{\emph{Proceedings of ACM/IEEE Joint Conference on Digital Libraries (JCDL)}}. \bibinfo{publisher}{IEEE Press}, \bibinfo{address}{Santa Fe, New Mexico, USA}, \bibinfo{pages}{102--106}.
\newblock
\urldef\tempurl%
\url{https://doi.org/10.1109/JCDL57899.2023.00025}
\showDOI{\tempurl}


\bibitem[Chang et~al\mbox{.}(2017)]%
        {chang2017}
\bibfield{author}{\bibinfo{person}{Li Chang}, \bibinfo{person}{Hsu-Chun Hsiao}, \bibinfo{person}{Wei Jeng}, \bibinfo{person}{Tiffany Hyun-Jin Kim}, {and} \bibinfo{person}{Wei-Hsi Lin}.} \bibinfo{year}{2017}\natexlab{}.
\newblock \showarticletitle{Security Implications of Redirection Trail in Popular Websites Worldwide}. In \bibinfo{booktitle}{\emph{Proceedings of the 26th International Conference on World Wide Web}} (Perth, Australia) \emph{(\bibinfo{series}{WWW '17})}. \bibinfo{publisher}{International World Wide Web Conferences Steering Committee}, \bibinfo{address}{Republic and Canton of Geneva, CHE}, \bibinfo{pages}{1491--1500}.
\newblock
\showISBNx{9781450349130}
\urldef\tempurl%
\url{https://doi.org/10.1145/3038912.3052698}
\showDOI{\tempurl}


\bibitem[Chapekis et~al\mbox{.}(2024)]%
        {pew2024}
\bibfield{author}{\bibinfo{person}{Athena Chapekis}, \bibinfo{person}{Samuel Bestvater}, \bibinfo{person}{Emma Remy}, {and} \bibinfo{person}{Gonzalo Rivero}.} \bibinfo{year}{2024}\natexlab{}.
\newblock \bibinfo{title}{When Online Content Disappears}.
\newblock
\newblock
\urldef\tempurl%
\url{https://www.pewresearch.org/data-labs/2024/05/17/when-online-content-disappears/}
\showURL{%
\tempurl}


\bibitem[Cho and Garcia-Molina(2000)]%
        {cho2000webevolution}
\bibfield{author}{\bibinfo{person}{Junghoo Cho} {and} \bibinfo{person}{Hector Garcia-Molina}.} \bibinfo{year}{2000}\natexlab{}.
\newblock \showarticletitle{The Evolution of the Web and Implications for an Incremental Crawler}. In \bibinfo{booktitle}{\emph{Proceedings of the 26th International Conference on Very Large Data Bases}} \emph{(\bibinfo{series}{VLDB '00})}. \bibinfo{publisher}{Morgan Kaufmann Publishers Inc.}, \bibinfo{address}{San Francisco, CA, USA}, \bibinfo{pages}{200–209}.
\newblock
\showISBNx{1558607153}


\bibitem[Contributors(2024)]%
        {Redirection_Codes}
\bibfield{author}{\bibinfo{person}{Baeldung Contributors}.} \bibinfo{year}{2024}\natexlab{}.
\newblock \bibinfo{title}{Redirection Status Codes - Complete Guide}.
\newblock
\newblock
\urldef\tempurl%
\url{https://www.baeldung.com/cs/redirection-status-codes}
\showURL{%
\tempurl}


\bibitem[Docs(2024a)]%
        {MDN_HTTP_Redirections}
\bibfield{author}{\bibinfo{person}{MDN~Web Docs}.} \bibinfo{year}{2024}\natexlab{a}.
\newblock \bibinfo{title}{HTTP Redirections}.
\newblock \bibinfo{howpublished}{Mozilla Developer Network}.
\newblock
\urldef\tempurl%
\url{https://developer.mozilla.org/en-US/docs/Web/HTTP/Redirections}
\showURL{%
\tempurl}


\bibitem[Docs(2024b)]%
        {MDNHttpStatus}
\bibfield{author}{\bibinfo{person}{MDN~Web Docs}.} \bibinfo{year}{2024}\natexlab{b}.
\newblock \bibinfo{title}{HTTP Status Codes}.
\newblock \bibinfo{howpublished}{Mozilla Developer Network}.
\newblock
\urldef\tempurl%
\url{https://developer.mozilla.org/en-US/docs/Web/HTTP/Status}
\showURL{%
\tempurl}


\bibitem[Fetterly et~al\mbox{.}(2003)]%
        {Fetterly2003}
\bibfield{author}{\bibinfo{person}{Dennis Fetterly}, \bibinfo{person}{Mark Manasse}, \bibinfo{person}{Marc Najork}, {and} \bibinfo{person}{Janet Wiener}.} \bibinfo{year}{2003}\natexlab{}.
\newblock \showarticletitle{A large-scale study of the evolution of web pages}. In \bibinfo{booktitle}{\emph{Proceedings of the 12th International Conference on World Wide Web}} (Budapest, Hungary) \emph{(\bibinfo{series}{WWW '03})}. \bibinfo{publisher}{Association for Computing Machinery}, \bibinfo{address}{New York, USA}, \bibinfo{pages}{669–678}.
\newblock
\showISBNx{1581136803}
\urldef\tempurl%
\url{https://doi.org/10.1145/775152.775246}
\showDOI{\tempurl}


\bibitem[Fielding and Reschke(2014)]%
        {rfc7231}
\bibfield{author}{\bibinfo{person}{Roy Fielding} {and} \bibinfo{person}{Julian Reschke}.} \bibinfo{year}{2014}\natexlab{}.
\newblock \bibinfo{title}{{Hypertext Transfer Protocol (HTTP/1.1): Semantics and Content, RFC 7231}}.
\newblock
\newblock
\urldef\tempurl%
\url{https://www.rfc-editor.org/rfc/rfc7231}
\showURL{%
\tempurl}


\bibitem[Garg(2024)]%
        {rickrolling_blog}
\bibfield{author}{\bibinfo{person}{Kritika Garg}.} \bibinfo{year}{2024}\natexlab{}.
\newblock \bibinfo{title}{Analyzing Redirects and Getting Rickrolled Along the Way}.
\newblock
\newblock
\urldef\tempurl%
\url{https://ws-dl.blogspot.com/2024/10/2024-10-22-analyzing-redirects-and.html}
\showURL{%
\tempurl}


\bibitem[Garg et~al\mbox{.}(2025)]%
        {Samplingtechreport:2025}
\bibfield{author}{\bibinfo{person}{Kritika Garg}, \bibinfo{person}{Sawood Alam}, \bibinfo{person}{Michele~C. Weigle}, {and} \bibinfo{person}{Michael~L. Nelson}.} \bibinfo{year}{2025}\natexlab{}.
\newblock \bibinfo{booktitle}{\emph{Longitudinal Sampling of URLs From the Wayback Machine}}.
\newblock \bibinfo{type}{{T}echnical {R}eport} arXiv:2507.14752. \bibinfo{institution}{arXiv}.
\newblock


\bibitem[Hall and Tiropanis(2012)]%
        {WebScience2012}
\bibfield{author}{\bibinfo{person}{Wendy Hall} {and} \bibinfo{person}{Thanassis Tiropanis}.} \bibinfo{year}{2012}\natexlab{}.
\newblock \showarticletitle{Web evolution and Web Science}.
\newblock \bibinfo{journal}{\emph{Computer Networks}} \bibinfo{volume}{56}, \bibinfo{number}{18} (\bibinfo{year}{2012}), \bibinfo{pages}{3859--3865}.
\newblock
\showISSN{1389-1286}
\urldef\tempurl%
\url{https://doi.org/10.1016/j.comnet.2012.10.004}
\showDOI{\tempurl}


\bibitem[Holzmann et~al\mbox{.}(2016)]%
        {holzmann2016dawn}
\bibfield{author}{\bibinfo{person}{Helge Holzmann}, \bibinfo{person}{Wolfgang Nejdl}, {and} \bibinfo{person}{Avishek Anand}.} \bibinfo{year}{2016}\natexlab{}.
\newblock \showarticletitle{The Dawn of Today's Popular Domains: A Study of the Archived German Web over 18 Years}. In \bibinfo{booktitle}{\emph{Proceedings of the 16th ACM/IEEE-CS Joint Conference on Digital Libraries (JCDL 2016)}}. \bibinfo{publisher}{ACM}, \bibinfo{address}{New Jersey, Newark, USA}, \bibinfo{pages}{73--82}.
\newblock
\urldef\tempurl%
\url{https://doi.org/10.1145/2910896.2910901}
\showDOI{\tempurl}


\bibitem[{International Organization for Standardization}(2017)]%
        {iso2017warc}
\bibfield{author}{\bibinfo{person}{{International Organization for Standardization}}.} \bibinfo{year}{2017}\natexlab{}.
\newblock \bibinfo{title}{Information and documentation — WARC file format}.
\newblock \bibinfo{howpublished}{\url{https://www.iso.org/standard/68004.html}}.
\newblock


\bibitem[{Internet Archive}(2023)]%
        {crawldata_IA}
\bibfield{author}{\bibinfo{person}{{Internet Archive}}.} \bibinfo{year}{2023}\natexlab{}.
\newblock \bibinfo{title}{{Not Your Parents' Web Dataset}}.
\newblock
\newblock
\urldef\tempurl%
\url{https://archive.org/details/not-your-parents-web}
\showURL{%
\tempurl}


\bibitem[Jack~Cushman(2017)]%
        {cushman:2017}
\bibfield{author}{\bibinfo{person}{Ilya~Kreymer Jack~Cushman}.} \bibinfo{year}{2017}\natexlab{}.
\newblock \bibinfo{title}{Thinking like a hacker: Security Considerations for High-Fidelity Web Archives}.
\newblock
\newblock
\urldef\tempurl%
\url{http://labs.rhizome.org/presentations/security.html}
\showURL{%
\tempurl}


\bibitem[Jones et~al\mbox{.}(2016)]%
        {jones2016scholarly}
\bibfield{author}{\bibinfo{person}{Shawn~M. Jones}, \bibinfo{person}{Herbert~Van de Sompel}, \bibinfo{person}{Harihar Shankar}, \bibinfo{person}{Martin Klein}, \bibinfo{person}{Richard Tobin}, {and} \bibinfo{person}{Claire Grover}.} \bibinfo{year}{2016}\natexlab{}.
\newblock \showarticletitle{Scholarly Context Adrift: Three out of Four URI References Lead to Changed Content}.
\newblock \bibinfo{journal}{\emph{PLOS ONE}} \bibinfo{volume}{11}, \bibinfo{number}{12} (\bibinfo{year}{2016}), \bibinfo{pages}{e0167475}.
\newblock
\urldef\tempurl%
\url{https://doi.org/10.1371/journal.pone.0167475}
\showDOI{\tempurl}


\bibitem[Kelly et~al\mbox{.}(2017a)]%
        {Mkelly:URIcanonicalization}
\bibfield{author}{\bibinfo{person}{Mat Kelly}, \bibinfo{person}{Lulwah~M. Alkwai}, \bibinfo{person}{Sawood Alam}, \bibinfo{person}{Michael~L. Nelson}, \bibinfo{person}{Michele~C. Weigle}, {and} \bibinfo{person}{Herbert Van~de Sompel}.} \bibinfo{year}{2017}\natexlab{a}.
\newblock \showarticletitle{Impact of URI Canonicalization on Memento Count}. In \bibinfo{booktitle}{\emph{Proceedings of the 17th ACM/IEEE Joint Conference on Digital Libraries}} \emph{(\bibinfo{series}{JCDL '17})}. \bibinfo{publisher}{IEEE Press}, \bibinfo{address}{Toronto, Ontario, Canada}, \bibinfo{pages}{303--304}.
\newblock
\showISBNx{9781538638613}


\bibitem[Kelly et~al\mbox{.}(2017b)]%
        {Mkelly:URIcanonicalization_techreport}
\bibfield{author}{\bibinfo{person}{Mat Kelly}, \bibinfo{person}{Lulwah~M. Alkwai}, \bibinfo{person}{Sawood Alam}, \bibinfo{person}{Michael~L. Nelson}, \bibinfo{person}{Michele~C. Weigle}, {and} \bibinfo{person}{Herbert Van~de Sompel}.} \bibinfo{year}{2017}\natexlab{b}.
\newblock \bibinfo{booktitle}{\emph{Impact of URI Canonicalization on Memento Count}}.
\newblock \bibinfo{type}{{T}echnical {R}eport} arXiv:1703.03302. \bibinfo{institution}{arXiv}.
\newblock


\bibitem[Klein et~al\mbox{.}(2014)]%
        {klein2014scholarly}
\bibfield{author}{\bibinfo{person}{Martin Klein}, \bibinfo{person}{Herbert~Van de Sompel}, \bibinfo{person}{Robert Sanderson}, \bibinfo{person}{Harihar Shankar}, \bibinfo{person}{Lyudmila Balakireva}, \bibinfo{person}{Ke Zhou}, {and} \bibinfo{person}{Richard Tobin}.} \bibinfo{year}{2014}\natexlab{}.
\newblock \showarticletitle{Scholarly Context Not Found: One in Five Articles Suffers from Reference Rot}.
\newblock \bibinfo{journal}{\emph{PLOS ONE}} \bibinfo{volume}{9}, \bibinfo{number}{12} (\bibinfo{year}{2014}), \bibinfo{pages}{e115253}.
\newblock
\urldef\tempurl%
\url{https://doi.org/10.1371/journal.pone.0115253}
\showDOI{\tempurl}


\bibitem[Kline et~al\mbox{.}(2019)]%
        {Kline2019}
\bibfield{author}{\bibinfo{person}{Jeffery Kline}, \bibinfo{person}{Edward Oakes}, {and} \bibinfo{person}{Paul Barford}.} \bibinfo{year}{2019}\natexlab{}.
\newblock \showarticletitle{A URL-based Analysis of WWW Structure and Dynamics}. In \bibinfo{booktitle}{\emph{Proceedings of the Network Traffic Measurement and Analysis Conference (TMA)}}. \bibinfo{publisher}{IEEE Press}, \bibinfo{address}{Paris, France}, \bibinfo{pages}{81--800}.
\newblock
\urldef\tempurl%
\url{https://doi.org/10.23919/TMA.2019.8784665}
\showDOI{\tempurl}


\bibitem[Koehler(1999)]%
        {koehler1999webpageconstancy}
\bibfield{author}{\bibinfo{person}{Wallace Koehler}.} \bibinfo{year}{1999}\natexlab{}.
\newblock \showarticletitle{An Analysis of Web Page and Web Site Constancy and Permanence}.
\newblock \bibinfo{journal}{\emph{Journal of the American Society for Information Science}} \bibinfo{volume}{50}, \bibinfo{number}{2} (\bibinfo{year}{1999}), \bibinfo{pages}{162--180}.
\newblock
\urldef\tempurl%
\url{https://doi.org/10.1002/(SICI)1097-4571(1999)50:2<162::AID-ASI7>3.0.CO;2-B}
\showDOI{\tempurl}


\bibitem[Koop et~al\mbox{.}(2020)]%
        {koop2020}
\bibfield{author}{\bibinfo{person}{Martin Koop}, \bibinfo{person}{Erik Tews}, {and} \bibinfo{person}{Stefan Katzenbeisser}.} \bibinfo{year}{2020}\natexlab{}.
\newblock \showarticletitle{In-depth Evaluation of Redirect Tracking and Link Usage}.
\newblock \bibinfo{journal}{\emph{Proceedings on Privacy Enhancing Technologies}} \bibinfo{volume}{2020}, \bibinfo{number}{4} (\bibinfo{year}{2020}), \bibinfo{pages}{394--413}.
\newblock
\showISSN{2299-0984}
\urldef\tempurl%
\url{https://doi.org/10.2478/popets-2020-0079}
\showDOI{\tempurl}


\bibitem[Kurkowski(2024)]%
        {TldExtract}
\bibfield{author}{\bibinfo{person}{John Kurkowski}.} \bibinfo{year}{2024}\natexlab{}.
\newblock \bibinfo{title}{tldextract}.
\newblock
\newblock
\urldef\tempurl%
\url{https://pypi.org/project/tldextract/}
\showURL{%
\tempurl}
\newblock
\shownote{Python package for extracting Top-Level Domain (TLD) from URLs}.


\bibitem[Lee et~al\mbox{.}(2009)]%
        {Lee2009}
\bibfield{author}{\bibinfo{person}{Taehyung Lee}, \bibinfo{person}{Jinil Kim}, \bibinfo{person}{Jin~Wook Kim}, \bibinfo{person}{Sung-Ryul Kim}, {and} \bibinfo{person}{Kunsoo Park}.} \bibinfo{year}{2009}\natexlab{}.
\newblock \showarticletitle{Detecting Soft Errors by Redirection Classification}. In \bibinfo{booktitle}{\emph{Proceedings of the 18th International Conference on World Wide Web}} (Madrid, Spain) \emph{(\bibinfo{series}{WWW '09})}. \bibinfo{publisher}{Association for Computing Machinery}, \bibinfo{address}{New York, USA}, \bibinfo{pages}{1119--1120}.
\newblock
\showISBNx{9781605584874}
\urldef\tempurl%
\url{https://doi.org/10.1145/1526709.1526886}
\showDOI{\tempurl}


\bibitem[Leitschuh(2019)]%
        {Leitschuh:2019}
\bibfield{author}{\bibinfo{person}{Jonathan Leitschuh}.} \bibinfo{year}{2019}\natexlab{}.
\newblock \bibinfo{title}{{Zoom Zero Day: 4+ Million Webcams \& maybe an RCE? Just get them to visit your website!}}
\newblock
\newblock
\urldef\tempurl%
\url{https://infosecwriteups.com/zoom-zero-day-4-million-webcams-maybe-an-rce-just-get-them-to-visit-your-website-ac75c83f4ef5}
\showURL{%
\tempurl}


\bibitem[{Moz Contributors}(2024)]%
        {Moz_Canonicalization}
\bibfield{author}{\bibinfo{person}{{Moz Contributors}}.} \bibinfo{year}{2024}\natexlab{}.
\newblock \bibinfo{title}{Canonicalization}.
\newblock
\newblock
\urldef\tempurl%
\url{https://moz.com/learn/seo/canonicalization}
\showURL{%
\tempurl}


\bibitem[Nelson(2021)]%
        {nypwblog2021}
\bibfield{author}{\bibinfo{person}{Michael~L. Nelson}.} \bibinfo{year}{2021}\natexlab{}.
\newblock \bibinfo{title}{{Not Your Parents’ Web: The Scope and Archiving of the Modern Web}}.
\newblock
\newblock
\urldef\tempurl%
\url{https://ws-dl.blogspot.com/2021/10/2021-10-20-not-your-parents-web-scope.html}
\showURL{%
\tempurl}


\bibitem[Ohye and Kupke(2012)]%
        {rfc6596}
\bibfield{author}{\bibinfo{person}{Maile Ohye} {and} \bibinfo{person}{Joachim Kupke}.} \bibinfo{year}{2012}\natexlab{}.
\newblock \bibinfo{title}{{The Canonical Link Relation}}.
\newblock
\newblock
\urldef\tempurl%
\url{https://www.rfc-editor.org/info/rfc6596}
\showURL{%
\tempurl}


\bibitem[{Reddit users}(2022)]%
        {reddit_pingerpl}
\bibfield{author}{\bibinfo{person}{{Reddit users}}.} \bibinfo{year}{2022}\natexlab{}.
\newblock \bibinfo{title}{Pinger.pl: A Polish Personal Blogging Platform - Discussion on ArchiveTeam}.
\newblock
\newblock
\urldef\tempurl%
\url{https://www.reddit.com/r/Archiveteam/comments/scp03k/pingerpl_a_polish_personal_blogging_platform/}
\showURL{%
\tempurl}


\bibitem[Sharma et~al\mbox{.}(2022)]%
        {Sharma2022}
\bibfield{author}{\bibinfo{person}{Simple Sharma}, \bibinfo{person}{Supriya~P. Panda}, {and} \bibinfo{person}{Seema Verma}.} \bibinfo{year}{2022}\natexlab{}.
\newblock \showarticletitle{Role and Analysis of Various SEO Strategies to Improve Website Ranking}. In \bibinfo{booktitle}{\emph{Proceedings of the International Conference on Machine Learning, Big Data, Cloud and Parallel Computing (COM-IT-CON)}}, Vol.~\bibinfo{volume}{1}. \bibinfo{publisher}{IEEE Press}, \bibinfo{address}{Faridabad, India}, \bibinfo{pages}{639--648}.
\newblock
\urldef\tempurl%
\url{https://doi.org/10.1109/COM-IT-CON54601.2022.9850597}
\showDOI{\tempurl}


\bibitem[Singh et~al\mbox{.}(2020)]%
        {singh2020}
\bibfield{author}{\bibinfo{person}{Kushagra Singh}, \bibinfo{person}{Gurshabad Grover}, {and} \bibinfo{person}{Varun Bansal}.} \bibinfo{year}{2020}\natexlab{}.
\newblock \showarticletitle{How India Censors the Web}. In \bibinfo{booktitle}{\emph{Proceedings of the 12th ACM Conference on Web Science}} (Southampton, United Kingdom) \emph{(\bibinfo{series}{WebSci '20})}. \bibinfo{publisher}{Association for Computing Machinery}, \bibinfo{address}{New York, USA}, \bibinfo{pages}{21--28}.
\newblock
\showISBNx{9781450379892}
\urldef\tempurl%
\url{https://doi.org/10.1145/3394231.3397891}
\showDOI{\tempurl}


\bibitem[Software et~al\mbox{.}(2021)]%
        {PywbIndexing}
\bibfield{author}{\bibinfo{person}{Webrecorder Software}, \bibinfo{person}{Rhizome}, {and} \bibinfo{person}{Contributors}.} \bibinfo{year}{2014--2021}\natexlab{}.
\newblock \bibinfo{title}{Zipnum Sharded Index}.
\newblock \bibinfo{howpublished}{pywb 2.7 Documentation}.
\newblock
\urldef\tempurl%
\url{https://pywb.readthedocs.io/en/latest/manual/indexing.html#zipnum-sharded-index}
\showURL{%
\tempurl}


\bibitem[Southern(2020)]%
        {MuellerRedirectAdvice}
\bibfield{author}{\bibinfo{person}{Matt~G. Southern}.} \bibinfo{year}{2020}\natexlab{}.
\newblock \bibinfo{title}{Google Recommends Less Than 5 Hops Per Redirect Chain}.
\newblock \bibinfo{howpublished}{Search Engine Journal}.
\newblock
\urldef\tempurl%
\url{https://www.searchenginejournal.com/googles-john-mueller-recommends-less-than-5-hops-per-redirect-chain/344664/}
\showURL{%
\tempurl}


\bibitem[Thompson(2024)]%
        {thompson2024}
\bibfield{author}{\bibinfo{person}{Henry~S. Thompson}.} \bibinfo{year}{2024}\natexlab{}.
\newblock \showarticletitle{Improved Methodology for Longitudinal Web Analytics Using Common Crawl}. In \bibinfo{booktitle}{\emph{Proceedings of the 16th ACM Web Science Conference}} (Stuttgart, Germany) \emph{(\bibinfo{series}{WebSci '24})}. \bibinfo{publisher}{Association for Computing Machinery}, \bibinfo{address}{New York, USA}, \bibinfo{pages}{59--69}.
\newblock
\showISBNx{9798400703348}
\urldef\tempurl%
\url{https://doi.org/10.1145/3614419.3644018}
\showDOI{\tempurl}


\bibitem[{Van de Sompel} et~al\mbox{.}(2013)]%
        {memento:rfc}
\bibfield{author}{\bibinfo{person}{Herbert {Van de Sompel}}, \bibinfo{person}{Michael~L. Nelson}, {and} \bibinfo{person}{Robert Sanderson}.} \bibinfo{year}{2013}\natexlab{}.
\newblock \bibinfo{title}{{HTTP framework for time-based access to resource states -- Memento, Internet RFC 7089}}.
\newblock \bibinfo{howpublished}{http://tools.ietf.org/html/rfc7089}.
\newblock


\bibitem[Wang et~al\mbox{.}(2005)]%
        {Wang2005}
\bibfield{author}{\bibinfo{person}{Xiaozhe Wang}, \bibinfo{person}{Ajith Abraham}, {and} \bibinfo{person}{Kate~A. Smith}.} \bibinfo{year}{2005}\natexlab{}.
\newblock \showarticletitle{Intelligent Web Traffic Mining and Analysis}.
\newblock \bibinfo{journal}{\emph{The Journal of Network and Computer Applications}} \bibinfo{volume}{28}, \bibinfo{number}{2} (\bibinfo{date}{April} \bibinfo{year}{2005}), \bibinfo{pages}{147--165}.
\newblock
\showISSN{1084-8045}
\urldef\tempurl%
\url{https://doi.org/10.1016/j.jnca.2004.01.006}
\showDOI{\tempurl}


\bibitem[Weigle(2024)]%
        {nypwfindingsblog2024}
\bibfield{author}{\bibinfo{person}{Michele~C. Weigle}.} \bibinfo{year}{2024}\natexlab{}.
\newblock \bibinfo{title}{Some URLs are Immortal, Most are Not}.
\newblock
\newblock
\urldef\tempurl%
\url{https://ws-dl.blogspot.com/2024/09/2024-09-20-some-urls-are-immortal-most.html}
\showURL{%
\tempurl}


\bibitem[{World Wide Web Consortium (W3C)}(2008)]%
        {w3c:metarefresh}
\bibfield{author}{\bibinfo{person}{{World Wide Web Consortium (W3C)}}.} \bibinfo{year}{2008}\natexlab{}.
\newblock \bibinfo{title}{{H76: Using the title attribute of the iframe element}}.
\newblock
\newblock
\urldef\tempurl%
\url{https://www.w3.org/TR/WCAG20-TECHS/H76.html}
\showURL{%
\tempurl}


\bibitem[Wu(2011)]%
        {Wu2011SearchEO}
\bibfield{author}{\bibinfo{person}{Huanwei Wu}.} \bibinfo{year}{2011}\natexlab{}.
\newblock \showarticletitle{Search Engine Optimization of E-Commerce Websites}. In \bibinfo{booktitle}{\emph{Proceedings of the International Conference on Management and Service Science}}. \bibinfo{publisher}{IEEE}, \bibinfo{address}{Wuhan, China}, \bibinfo{pages}{1--3}.
\newblock
\urldef\tempurl%
\url{https://api.semanticscholar.org/CorpusID:35218016}
\showURL{%
\tempurl}


\bibitem[Zhang and Cabage(2017)]%
        {Zhang2017}
\bibfield{author}{\bibinfo{person}{Sonya Zhang} {and} \bibinfo{person}{Neal Cabage}.} \bibinfo{year}{2017}\natexlab{}.
\newblock \showarticletitle{Search Engine Optimization: Comparison of Link Building and Social Sharing}.
\newblock \bibinfo{journal}{\emph{Journal of Computer Information Systems}}  \bibinfo{volume}{57} (\bibinfo{year}{2017}), \bibinfo{pages}{148 -- 159}.
\newblock
\urldef\tempurl%
\url{https://api.semanticscholar.org/CorpusID:63144097}
\showURL{%
\tempurl}


\bibitem[Zittrain et~al\mbox{.}(2021)]%
        {zittrain2021paper}
\bibfield{author}{\bibinfo{person}{Jonathan~L. Zittrain}, \bibinfo{person}{John Bowers}, {and} \bibinfo{person}{Clare Stanton}.} \bibinfo{year}{2021}\natexlab{}.
\newblock \showarticletitle{The Paper of Record Meets an Ephemeral Web: An Examination of Linkrot and Content Drift within The New York Times}.
\newblock \bibinfo{journal}{\emph{SSRN Electronic Journal}} (\bibinfo{year}{2021}), \bibinfo{pages}{1--13}.
\newblock
\urldef\tempurl%
\url{https://doi.org/10.2139/ssrn.3833133}
\showDOI{\tempurl}


\end{thebibliography}

\end{document}